\documentclass[useAMS,usenatbib,usegraphicx]{mn2e}

\usepackage{subfigure}

\title[Microlensing of a Reverberating Broad Line Region
]{Gravitational Microlensing of a Reverberating Quasar Broad Line Region
-- I. Method and Qualitative Results}
\author[H. Garsden et al.]{H. Garsden, N. F. Bate and G. F. Lewis\thanks{E-mail:
hgar7294@uni.sydney.edu.au (HG); nbate@sydney.usyd.edu.au (NFB); geraint.lewis@sydney.edu.au (GFL). Research undertaken as part of the Commonwealth Cosmology Initiative (CCI: www.thecci.org), an international collaboration supported by the Australian Research Council.}\\
Sydney Institute for Astronomy, School of Physics, A28, The University of Sydney,
NSW, 2006, Australia}
\begin{document}

\date{Accepted --. Received --; in original form 14 Jun 2011}

\pagerange{\pageref{firstpage}--\pageref{lastpage}} \pubyear{2011}

\maketitle

\label{firstpage}

\begin{abstract}
The kinematics and morphology of the broad emission line region (BELR) of quasars are  the subject of significant debate. The two leading methods for constraining BELR properties are microlensing and reverberation mapping. Here we combine these two methods with a study of   the microlensing behaviour of the BELR in Q2237+0305, as a change in continuum emission (a ``flare'') passes through it. Beginning with some generic models of the BELR -- sphere, bicones, disk -- we  slice in velocity and time  to produce brightness profiles of the BELR over the duration of the flare. These are  numerically microlensed to determine whether microlensing of reverberation mapping  provides new information about the properties of BELRs. We describe our method and show images of the models as they are flaring, and the unlensed and lensed spectra that are produced. Qualitative results and a discussion of the spectra are given in this paper, highlighting some  effects that could be observed. Our conclusion  is that the influence of microlensing,  while not strong, can produce significant observable effects that will help in differentiating  the properties of BELRs.
\end{abstract}

\begin{keywords}  
gravitational lensing: micro -- quasars: individual: Q2237+0305 -- galaxies: structure -- broad line region -- methods: numerical
\end{keywords}

\label{sec:micro_sims}

\begin{figure}
  \includegraphics[scale=0.455]{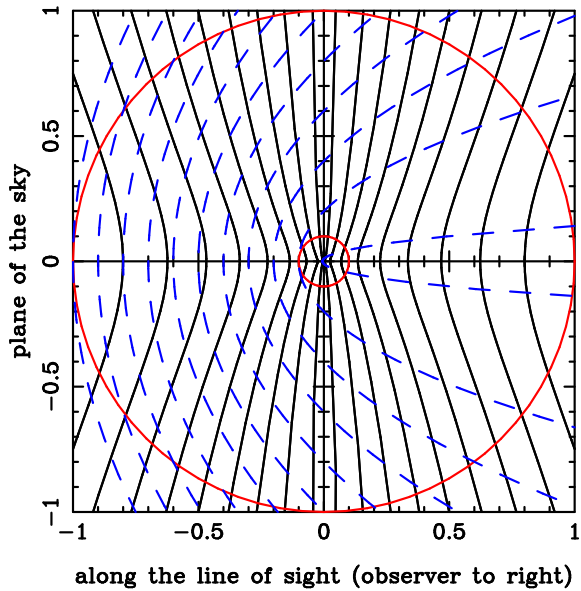}
  \caption{A projection of lines of constant time delay (dashed blue) and line-of-sight velocity (solid black) in a broad-line region during a flare, with the observer to the right of the plot. The constant time delay surfaces progress from right to left over time. The radial  velocities increase out from the origin, via Equation \ref{eqn:velocity} with $p = 0.5$.  The red lines superimpose the inner and outer radius of a circular BELR. The axes are in units normalized to a radius of 1.}
  \label{constant_vlos}
\end{figure}

\section{Introduction}
The geometry and kinematics of quasar broad emission line regions (BELRs) are currently poorly constrained. This region is too small to be resolved with current generation telescopes,  so we must rely on indirect techniques to probe its structure. The most prominent technique  in use is reverberation mapping (\citealt{blandford+82}; \citealt{peterson93}), however gravitational microlensing is a promising alternative (e.g. \citealt{nemiroff88}; \citealt{schneider+90}; \citealt{lewis+98}; \citealt{abajas+02};  \citealt{lewis+04}; \citealt{abajas+07}; \citealt{odowd+11}; \citealt{sluse+11}). In this paper, we discuss combining the two techniques, and conduct a numerical analysis of reverberation mapping in a gravitationally microlensed quasar, using several  generic BELR models.

Perhaps the most prominent feature of quasar spectra are their broad emission lines. The line broadening is thought to be caused by the Doppler motions of the BELR gas, sometimes giving rise to line widths of up to tens of thousands of kilometres. The absence of [OIII]$\lambda\lambda4364$, $4959$, $5007$ lines in quasar spectra, and the presence of the strongly non-permitted CIII]$\lambda1909$ line, allow us to infer electron densities in the BELR that suggest filamentary or clumpy structure. It is therefore natural to consider the BELR to consist of clouds of gas radiating at their thermal width, undergoing bulk motions large enough to produce the observed emission line widths.

Most of the information we have on the kinematics and structure of this region comes from reverberation mapping measurements. These experiments observe the response of the BELR to variations in the continuum flux from the active galactic nucleus (AGN). If there is a sudden increase in flux in the continuum, which we term a ``flare'', this  propagates through the BELR as an expanding shell and  is re-emitted by the material it passes through.  At the simplest level, reverberation mapping provides a measurement of the radius at which the emitting BELR gas is located. These measurements suggest stratification of the BELR, since higher ionisation lines such as CIV  are emitted at radii closer to the source than lower ionisation lines such as CIII] 
\citep{peterson97}. They also allow for the calculation of central black hole masses, under the assumption that virial motions dominate the BELR velocity structure (see for example \citealt{peterson+04}; \citealt{bentz+09}).

Recently, sampling rate and data quality have begun to reach a level where velocity-resolved time lags across broad emission lines can be measured (\citealt{horne+04}; \citealt{bentz+08}; \citealt{denney+09}; \citealt{bentz+10}). This was the original promise of the reverberation mapping technique: mapping the emissivity and velocity structure of the BELR. The results so far are ambiguous, indicating orbital, inflowing and outflowing bulk motion of the BELR gas (see especially \citealt{denney+09} and references therein). The interpretation of these signals remains problematic, but new methods, such as using Bayesian inference to iteratively converge on BELR parameters from light curves \citep{pancoast2011,brewer2011}, will help to remove some of the ambiguities.

So far, reverberation mapping experiments have been mostly limited to low redshift, low luminosity AGN. The one notable exception is S5 0836+71, a redshift $z=2.172$ quasar with a tentative detection of a reverberation mapping time delay of 595 days in the observer's frame \citep{kaspi+07}. This was the result of approximately 5 years ($\sim1.2$-$1.5$yr in the quasar rest frame) spectroscopic monitoring of 6 high-redshift objects. Amongst other complications, high luminosity AGN have slower and smaller continuum variations than their low luminosity counterparts.

A complementary technique for constraining the detailed kinematic and geometric structure of the BELR is gravitational microlensing. In this scenario, a background quasar is multiply imaged by a foreground galaxy. Stars and other compact matter in the lensing galaxy can (de)magnify different regions of the BELR. 
Gravitational microlensing as a probe of BELR structure was first discussed in \citet{nemiroff88}, where the effect of a single low mass star was considered on a range of kinematical models for the BELR. This work focused on broad emission line shapes. \citet{schneider+90} noted that line shape is a difficult diagnostic to use, since they can vary considerably between sources, and within sources over time. Instead, they suggested comparisons between images in the same lensing system, since the underlying line profile should be identical, but the microlensing-induced variations are uncorrelated
between images.

Subsequent theoretical analyses of the effect of microlensing on quasar BELRs were undertaken by \citet{abajas+02}, and \citet{lewis+04}. Both found that broad emission line profiles could be significantly affected by microlensing. \citet{lewis+04} noted that spectroscopic monitoring of lensed quasars is required for obtaining detailed morphology of the BELR. Until very recently, this sort of expensive observational campaign has not been feasible. However, newer techniques such as cheap, efficient photometric observations of reverberation mapping \citep{chelouche2011} have been proposed as an alternative.

Some applications of  gravitational microlensing analysis to actual BELR observational data have been made. \citet{lewis+98} demonstrated that the BELR of Q2237+0305 was undergoing microlensing through measurements of line equivalent widths, and \citet*{wayth+05} used IFU data to constrain the size of the CIII]/MgII broad emission line region in the same quasar,
finding it had a radius of $\sim0.06h^{1/2}_{70}$pc. 

More detailed spectroscopic analyses of broad emission lines in Q2237+0305 were undertaken by \citet{odowd+11} and \citet{sluse+11}.  \citet{odowd+11} detected a differential microlensing signal  across the CIII] emission in a single epoch of data. The data were compared to an outflowing and an orbital model of the BELR, and were found to favour the orbital model. \citet{sluse+11} instead focused on a variety of techniques for analysing the lines directly, such as multi-component decomposition. They measured the sizes of the CIII] and CIV emission regions, and found that each line was emitted from two spatially distinct regions: a compact high velocity component, and a larger low velocity component. The half-light radius of the CIV region was found to be $\sim 0.06$ pc.
These results were also consistent with a BELR dominated by virial motions.

SDSS J1004+4112 has also been the subject of several observations. \citet{richards2004} found enhancement of the blue wings of some emission lines in  SDSS J1004+4112 that were attributed to microlensing. The blue wing enhancement was also observed by \citet{gomez2006}, but the absence of other  microlensing-induced changes  suggested intrinsic variability may  be the cause. The latter has been supported by  \citet{lamer2006}, who suggest it provides a better explanation than microlensing for  differences in X-ray and UV variability.
On the other hand, \citet{ota2006} supports the microlensing hypothesis for X-ray flux anomalies and differences between X-ray and optical. \citet{abajas+07} attempted to reproduce   the blue wing enhancement numerically, using a   biconical BELR model, and found that it could reproduce the observed signal, albeit with a low probability. Clearly, there is more work to be done to confirm microlensing, and BELR microlensing, in this system.

A lensed quasar that does  show strong evidence of BELR microlensing is SDSS J0924+0219,
 based on flux anomalies, in both the continuum spectra and broad-emission lines, between the different  lensed images.  The observations were made by  \citet{keeton2006}  who derive a   BELR size of $\sim$ 9 light days in this quasar.

The above microlensing analyses all assumed that the BELR emission was constant and unvarying. In this paper, we relax this assumption and explore the possibility of using gravitationally lensed quasars to perform reverberation mapping experiments.
Using numerical microlensing techniques we model a flare passing through a BELR and observe the unlensed and lensed spectrum at several time intervals during the flare. We discuss reverberation and numerical microlensing in Section \ref{sec:background}. In Section \ref{sec:method}, we discuss the method used to obtain the data in this and subsequent papers. Results of the simulations are presented in Section \ref{sec:quals}. In Section \ref{sec:discuss} we discuss some of the effects produced by the microlensing of a flare. Our conclusions, and some implications for future work, are presented  in Section \ref{sec:conc}.

Throughout this paper, a cosmology with $H_0 = 70\rmn{km}\rmn{s}^{-1}\rmn{Mpc}^{-1}$, $\Omega_{m} = 0.3$ and $\Omega_{\Lambda} = 0.7$ is assumed.

\section{Background}
\label{sec:background}

\subsection{Broad emission-line region}

In line with previous BELR microlensing analyses (particularly \citealt{abajas+02} and \citealt{lewis+04}), we make use of the BELR models of \citet{robinson95} to specify properties of a BELR. These models are not intended to be physically realistic. Rather, they focus on a few key features of the BELR: the velocity distribution of the BELR clouds, their emissivity, and the BELR size and orientation.  

The BELR is assumed to consist of many clouds that re-radiate emission from the core. We adopt the following independent power-law relationships for the emissivity $\epsilon$ and the magnitude of the velocity $v$ of the emitting clouds in the BELR:

\begin{equation}
\epsilon = \epsilon_0\left(\frac{r}{r_0}\right)^\beta
\label{eqn:emiss}
\end{equation}
and

\begin{equation}
v = v_0\left(\frac{r}{r_0}\right)^p
\label{eqn:velocity}
\end{equation}
where $r_0$ is the inner radius of the cloud system. 

The efficiency with which the individual BELR clouds reprocess incoming radiation is assumed to be a smooth power-law function of radius. The volume density is also assumed to be a power-law function of radius. These two power-laws are
implicitly combined into  Equation \ref{eqn:emiss} so that, for a suitable choice of $\beta$ and $\epsilon_0$, the equation
can be used as the flux at a distance $r$ from the core. We assume that the number of clouds is large enough to approximate a smooth distribution. The clouds are assumed to emit isotropically, which implies that either the gas is optically thin or the BELR clouds are illuminated uniformly by diffuse ionising radiation. This assumption is almost certainly incorrect (see \citealt{robinson95} for details), however it is made here for the sake of simplicity.
The clouds are also assumed to re-radiate emission received only from the core, and not from other clouds. This is also incorrect since clouds may excite each other, but we also ignore that here.

The kinematic properties of the BELR are responsible for the line-broadening via Equation \ref{eqn:velocity}, and,  with the morphology and emission of the BELR, determine the spectrum. Different values of $p$ in Equation \ref{eqn:velocity} can considerably alter the  spectrum -- see \citet{abajas+02} for examples. The direction of the velocity may be inflowing, outflowing, or orbital. The non-linear nature of Equation \ref{eqn:velocity} also indicates that inflows and outflows are radially accelerating or decelerating.

\subsection{Reverberation mapping}
\label{sec:reverb}

A flare in the AGN continuum can be used to map the BELR as the flare reverberates through it 
\citep{kaspi+07}. Of particular significance is the time delay between the continuum flare and its appearance in and through the BELR, which allows the BELR size to be measured \citep{peterson93}. 
Suppose at time $\tau=0$ a flare is emitted from the centre of the BELR ($r=0$). It travels outward at the speed of light, causing the BELR gas to brighten as it passes. In spherical polar coordinates ($\phi$ is the zenith angle), the time at which a cloud is seen to flare by an observer is described by the following relationship:
\begin{equation}
\tau = \left(\frac{r}{c}\right)(1 - \rmn{cos}\theta\,\rmn{sin}\phi).
\end{equation}
Although the flare propagates radially outwards, the observer sees the BELR brighten initially along the line-of-sight. The flare then appears to expand to cover the BELR until it diminishes again to the line-of-sight. At each time the observer is viewing a paraboloid surface spreading away from the observer through the BELR \citep{horne+04}.  Figure \ref{constant_vlos} depicts some of these properties  with a side on view of a spherical  BELR with $p = 0.5$ in Equation \ref{eqn:velocity}, and the observer to the right. The red lines are the inner and outer radius of the BELR.  The blue lines are lines of constant time delay,  and they move from right to left over time. Also shown are lines of constant line-of-sight velocity (black), where the smallest velocity is at the vertical-axis, and increasing velocities to the left and right.

\subsection{Numerical microlensing}

When a quasar is gravitationally lensed by a foreground galaxy, multiple quasar images are produced due to the overall gravitational potential of the galaxy. These are magnified and distorted, and usually not resolvable into  substructure.  Modelling the location and magnifications of the  images  can be done analytically by assuming a smooth mass distribution for the lensing galaxy, which generates two key parameters for the mass distribution responsible for the images:  the convergence ($\kappa$),  and shear ($\gamma$). The convergence specifies the mass density in the vicinity of a light ray, the shear incorporates the effect of the overall mass distribution of the lensing galaxy. 

The quasar moves relative to the line-of-sight from the observer to the lensing galaxy, and as it moves the  structure within the galaxy -- stars, planets, clouds, etc. -- shifts relative to the light paths producing the quasar images. This can cause the images to fluctuate in magnification, an   effect that is called ``microlensing''. Microlensing has been observed in many lensed quasars 
\citep[e.g.][]{walsh1979,huchra1985,turner1989,myers1999,inada2005,abajas+07,kayo2010}. 

\begin{figure}
\includegraphics[scale=.445]{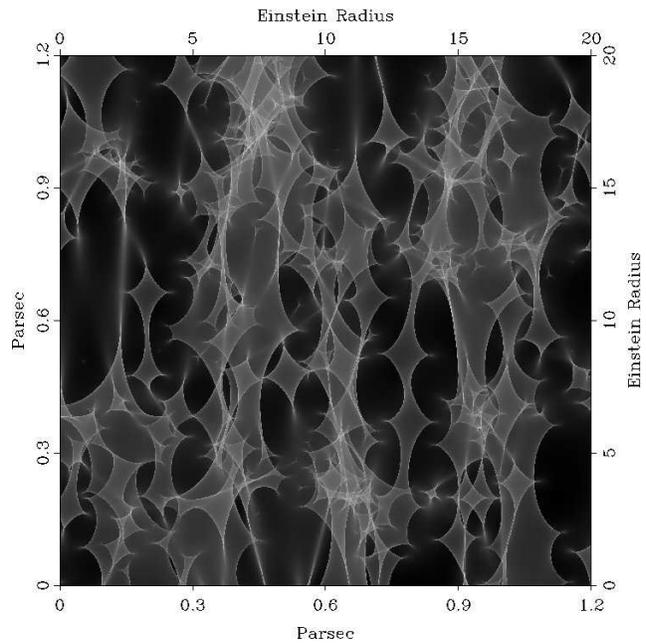}\caption{A magnification map  generated for image A in Q2237+0305, and used in this paper. The brightness at a location indicates the magnification, due to microlensing, of a pixel-sized source situated at that location. Bright areas indicate high magnification, dark areas indicate low magnification, relative to the mean magnification of image A. The map was generated using a convergence 
$\kappa = 0.36$ and shear $\gamma = 0.40$. The map covers a region in the source plane of $1.2 \times 1.2$ pc$^2$ ($20 \times 20 $ ER$^2$) at a resolution of 10000$^2$ pixels.}
\label{map}
\end{figure}

To model microlensing,  the substructure in the lens galaxy  must be taken into account, and since there are very many compact objects, microlensing is modelled numerically. Using the parameters above,
  compact objects and smooth matter are generated  and  light rays fired through them to calculate the deflections. It is computationally easier and more efficient to fire the rays from the observer through the lens to the plane of the source --  this is called ``inverse ray-tracing''.
The rays hit a pixelized source plane behind the galaxy  and are counted for each pixel;
the number of rays reaching a pixel is the change in magnification, due to microlensing, of a pixel-sized source at that location.
The result of inverse ray-tracing is a magnification map generated for each quasar image; Figure \ref{map} shows the map for image A in Q2237+0305 that is used in this work. The map represents an area of sky over which a pixel-sized quasar source may be located; the mean magnification of the map is the magnification of the quasar image. Bright areas are where the quasar will be magnified, dark areas are where it will be demagnified, relative to the mean. These regions are separated by lines  where the magnification becomes formally infinite, these are called ``caustics''.

Quasars  may be larger than a pixel. Extended sources such as these are modelled by convolving a source profile with the magnification map, producing another map representing the microlensing of the source. In microlensing, small sources exhibit the most variability in magnification, as large sources will blur the sharpness of the magnification map \citep{lewis2006a,bate2008a}. Sources of different shapes and sizes \citep{mortonson2005} may be magnified differently at different locations on the map because they lie over different regions of magnification. Similarly if the source is emitting different wavelengths from different locations, the spectrum may be altered  \citep{wambsganss1991,mosquera2009} because the locations within the source profile may be magnified differently.  This chromatic microlensing can be modelled by  convolving  a region of one wavelength  with the magnification map,  doing that for many regions, and reconstructing the microlensed spectrum. It will be used to study the effect of microlensing on our BELR models because the velocity varies by location over the BELR, so for certain locations on the map  chromatic microlensing will be present.

Sources  of similar size, or smaller, than the characteristic size scale used in microlensing -- the Einstein Radius (ER) --  will be more susceptible to microlensing \citep{wyithe2000}. The Einstein Radius ($\eta_0$) is derived from the mass of a single point lens and distances from the observer to the lens and to a point source:
\begin{equation}
\label{Eq:EinsteinRadius}
\eta_0=\sqrt{\frac{4 G M}{c^2} \frac{D_{os} D_{ls}}{D_{ol}}},
\end{equation}
 where $G$ is the gravitational constant, $M$ is the mass of the lens and $D_{xy}$ refers to the angular diameter distance between $x$ and $y$; the subscripts $s, l,$ and $o$ representing source, lens, and observer respectively. 
In Q2237+0305, with a distance to the lens of $z_L = 0.0394$, and to the source of $z_S = 1.695$, and for a point lens of 1 Solar mass (M$_\odot$), the Einstein Radius is 0.06 pc.

\section{Method}
\label{sec:method}
We use Q2237+0305 for our lensed system as it is one of the most monitored lensed quasars \citep{udalski2008}, and it is an attractive laboratory for microlensing studies because it has a large Einstein Radius and therefore high microlensing variability  \citep{Kayser:89}. While not  a target of study for reverberation mapping, as significant BELR variability has not been reported,  its BELR is one of the most studied \citep[eg.]{lewis+04,lewis2006a,wayth+05,odowd+11,sluse+11}, and will continue to be so \citep{mosquera2011}.  We use it for illustrative purposes here. We generate source models for the BELR in Q2237+0305 that are  sliced by time and velocity as a flare is propagating through the BELR. The slices are microlensed, and combined into a time-varying spectrum over the lifetime of the flare.

\subsection{Magnification Maps}

 For this paper we focus upon image A of Q2237; in a later contribution  we will expand this to consider all images in this system.
 The magnification map we have generated for  this image is shown in Figure \ref{map}. The convergence and shear values used are $\kappa = 0.36$, $\gamma = 0.4$  \citep{schmidt1998}.  We generate 1,575 objects with 1 M$_\odot$ masses for the lens, and no smooth matter, as the images are located in the bulge where  stellar matter dominates.  We use the inverse ray-tracing method developed by \citet{wambsganss1990,wambsganss1999} and \citet{garsden2010} to fire the rays.    
The width of the map is 
20 ER (1.2 pc) at a pixel resolution  of 10000$^2$ pixels, or 0.002 ER (0.00012 pc) per pixel. 

\subsection{Model profiles}

\begin{figure*}
\centering
\subfigure[Sphere]{\includegraphics[width=55.3mm]{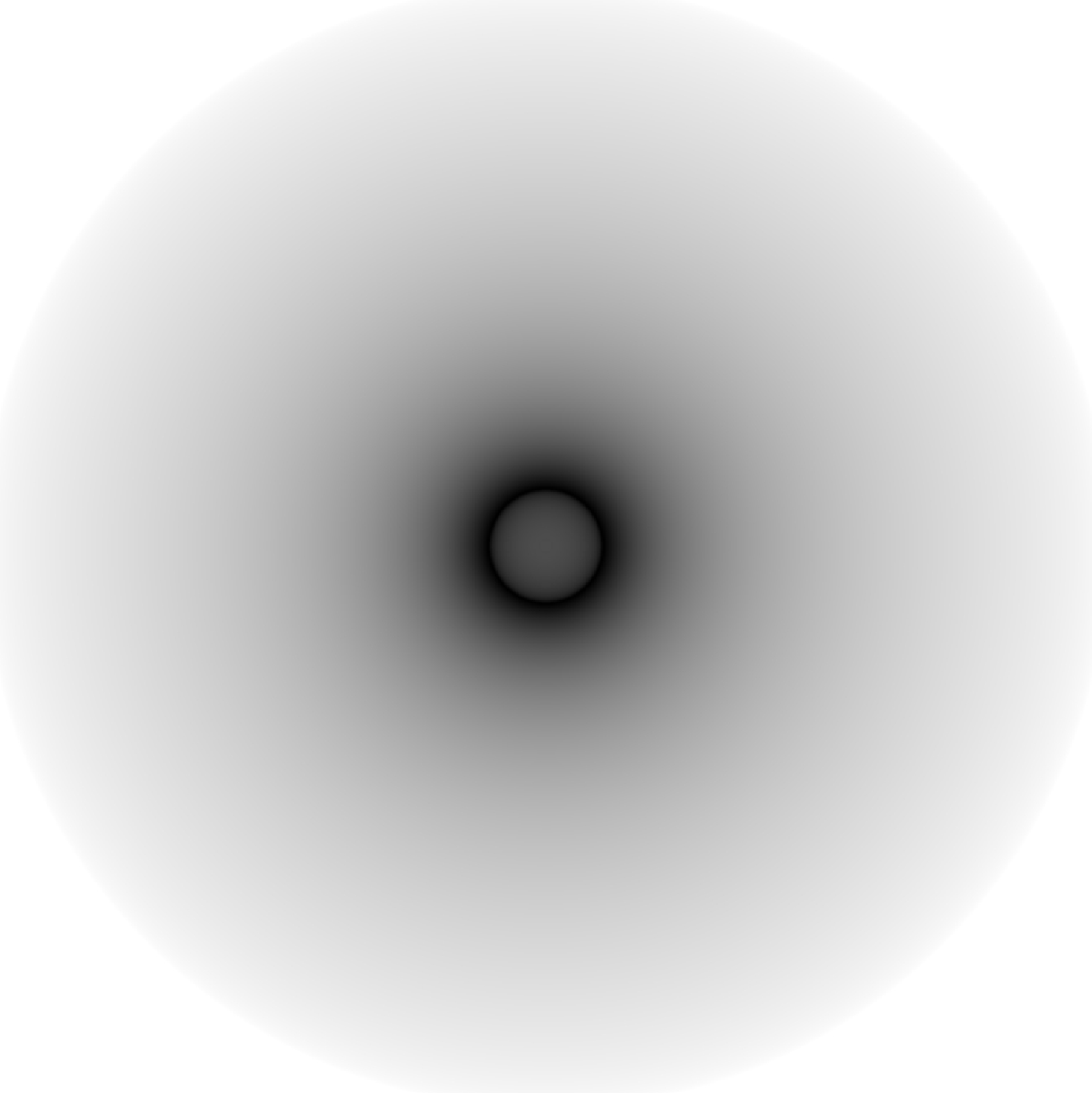}\label{sphere profile}}
\subfigure[Solid bicones, inclined 90$\,^\circ$, RA$=0\,^\circ$]{\includegraphics[width=55.3mm]{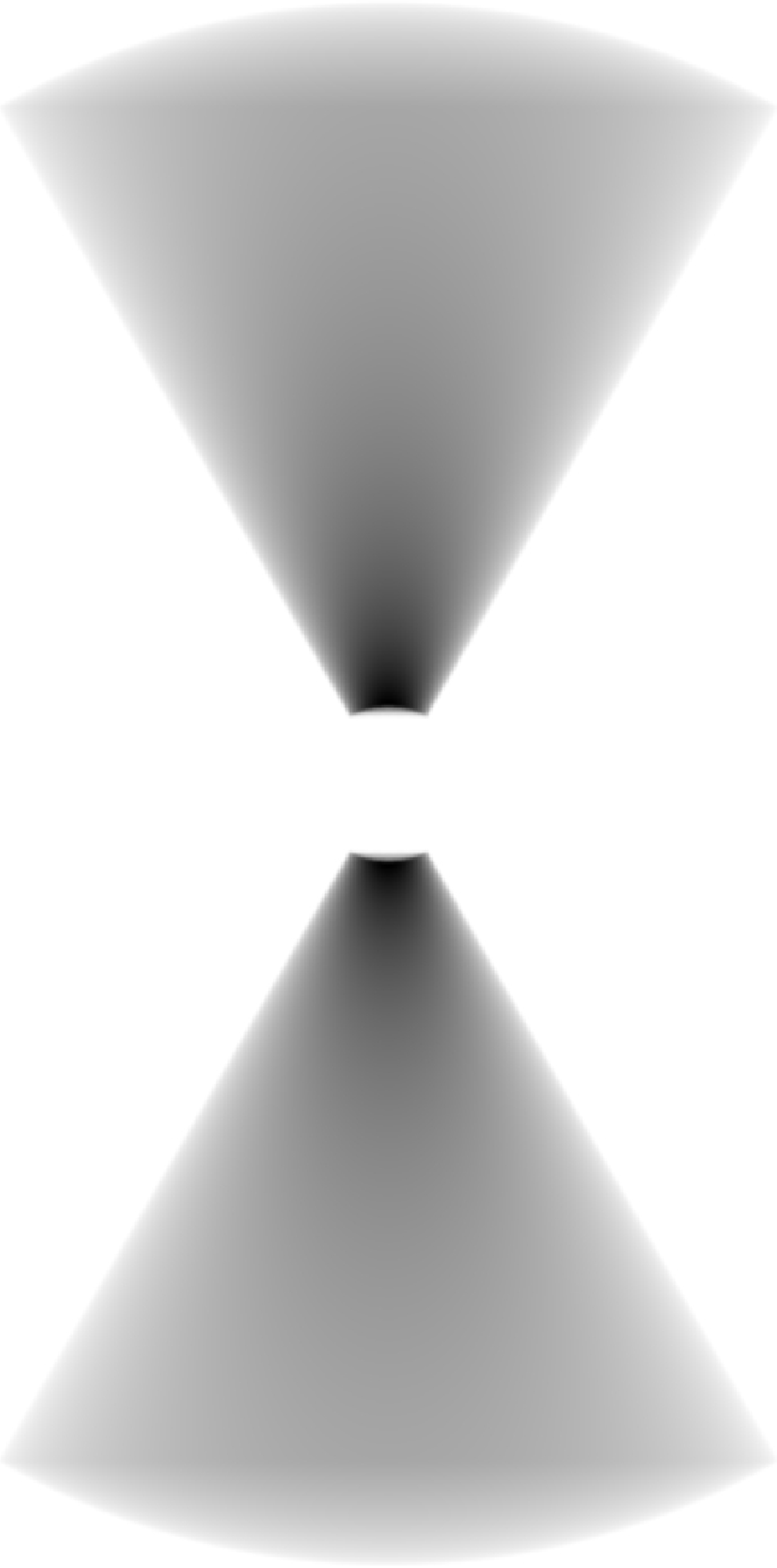}\label{bicone profile}}\\
\subfigure[Hollow bicones,  inclined 35$\,^\circ$, RA$=45\,^\circ$ ]{\includegraphics[width=55.3mm]{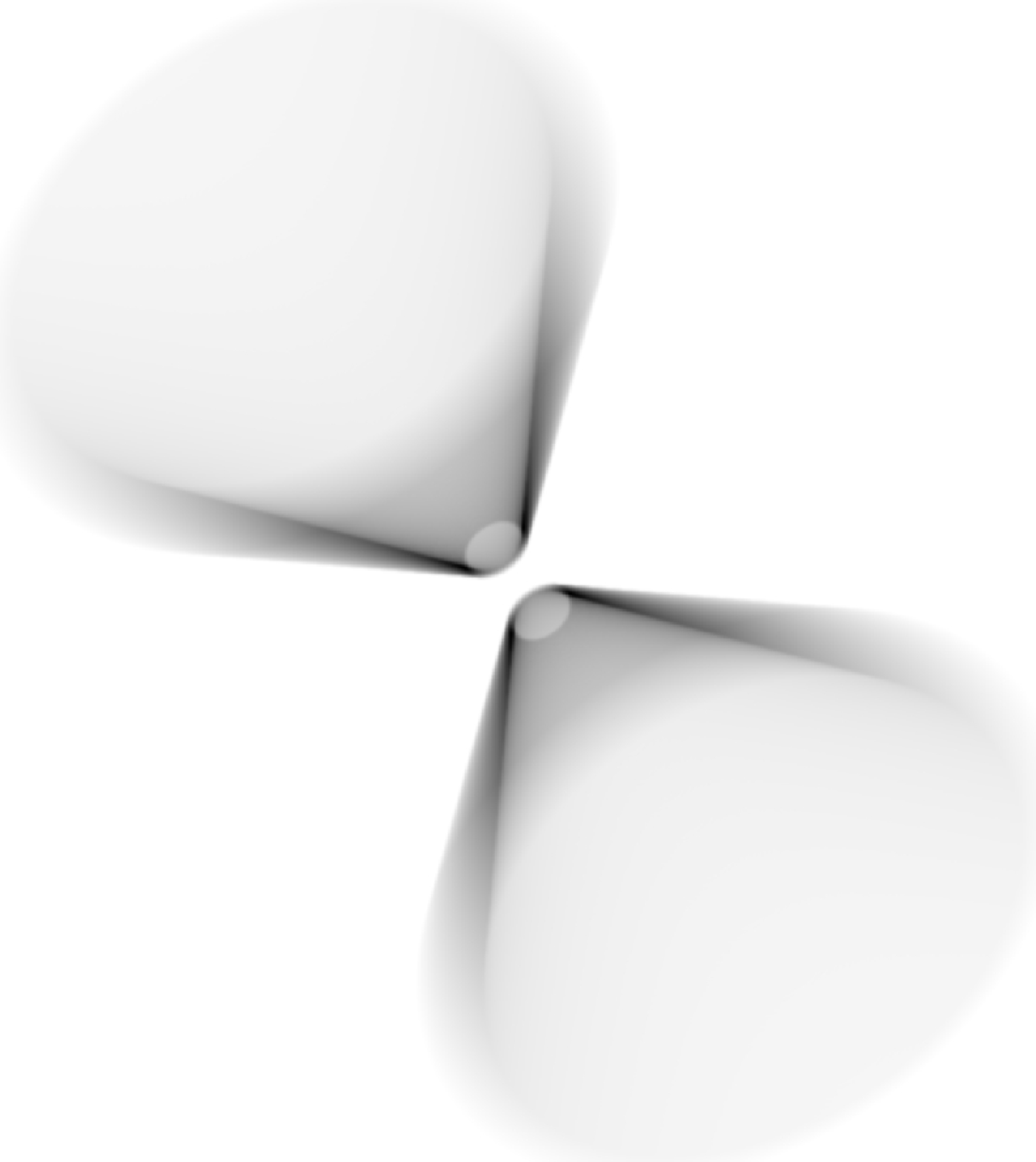}\label{hollow bicone profile}}
\subfigure[Disk, inclined 45$\,^\circ$, RA$=45\,^\circ$ ]{\includegraphics[width=55.3mm]{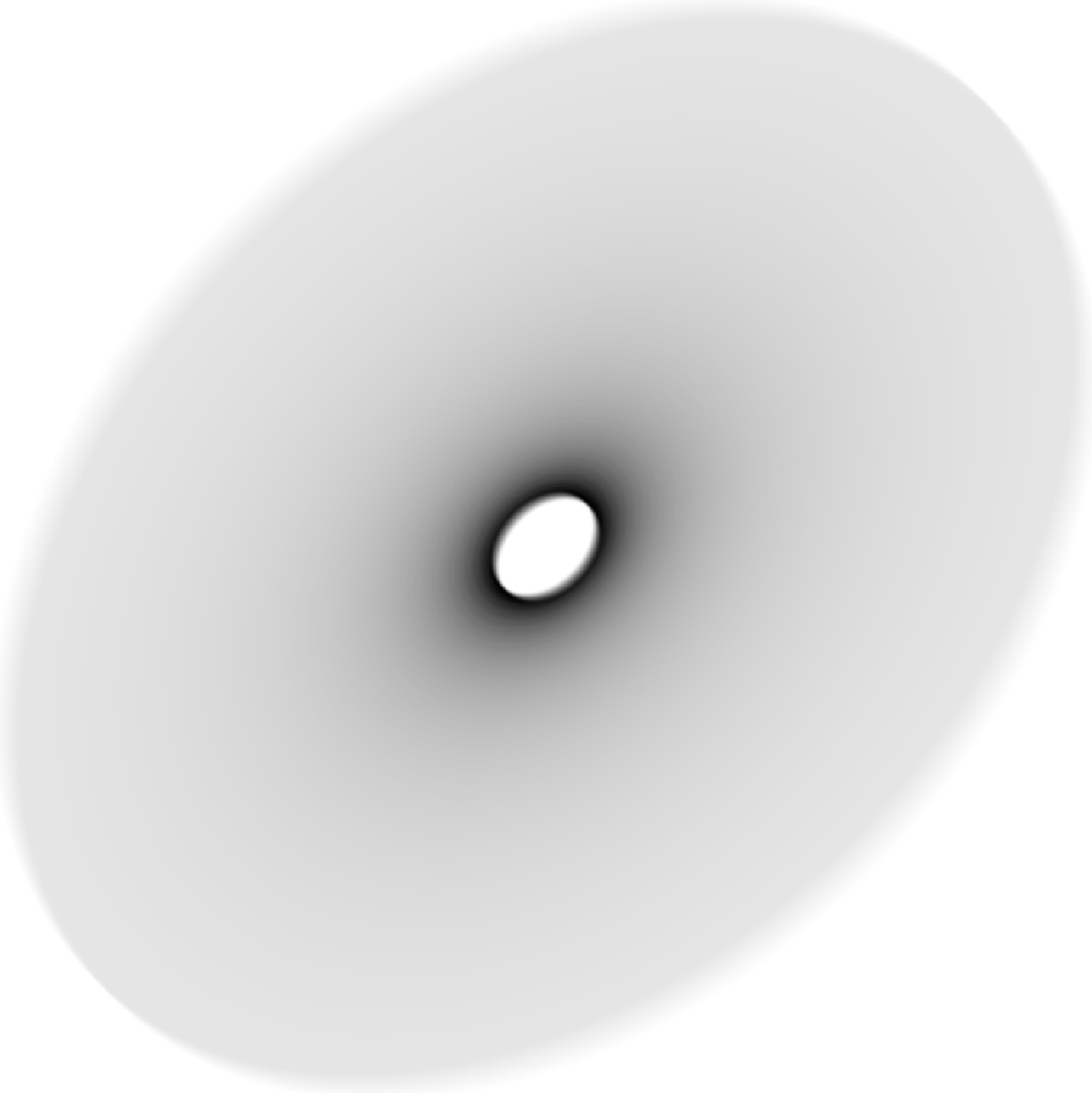}\label{disk profile}}
\caption{The source models used in this study. (a) is a sphere of uniformly distributed, transparent,  inflowing material. (b) consists of two cones, with an opening angle of 30$\,^\circ$, cut from the sphere in (a), and with the material  outflowing. 
(c) is created from (b) by hollowing out the cones and replacing the surface with a shell wall that has an opening angle of 
3$^\circ$. (d) is a  Keplerian disk of thickness 1\% of its radius. The bicones and disk models can be oriented
by inclining them in a certain direction. The orientation is indicated for each model. RA indicates an anti-clockwise rotation on the plane of the magnification map.}
\label{models}
\end{figure*}

We use four different source models, three taken from \citet{abajas+02}: the sphere, disk, and bicone.
There is also  increasing evidence for a disk-wind model \citep{elvis+00, peterson2006}, which adds a  funnel to the disk. For this paper and for simplicity we will implement a hollow bicone as  the funnel, and examine it separately from the disk. In future work a full ``disk-wind'' model can be implemented by combining the 
disk and the hollow bicones. 

We set  the size  of the models to be 1 ER = 0.06 pc, in line with sizes measured previously by \citet{wayth+05} and \citet{sluse+11}. The models are oriented and projected onto the sky to produce a source profile for microlensing.  All models have an inner 
radius of 10\% of the total radius, i.e. $r_0$ = 0.1 $r_{BELR}$ in Equation \ref{eqn:emiss}, where $r_{BELR}$ is the outer radius of the BELR. The BELR material is transparent, and at each point emits with flux given by Equation \ref{eqn:emiss}, with $\beta = -1.5$. The models are described below.

\begin{figure*}
\centering
\subfigure[Sphere]{\includegraphics[scale=0.38]{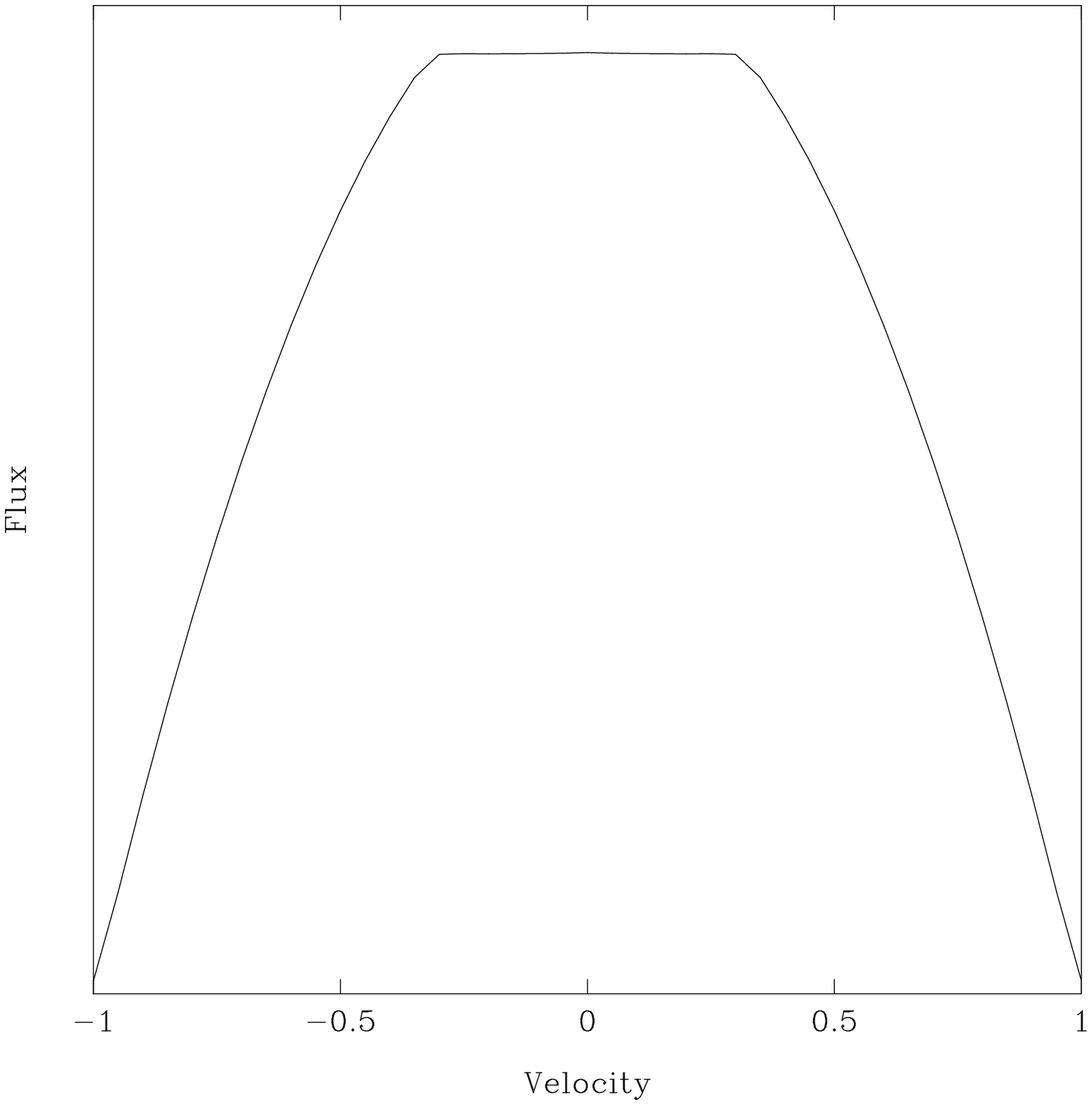}\label{sphere spectrum}}
\subfigure[Solid bicones, inclined 90$\,^\circ$, RA$=0\,^\circ$]{\includegraphics[scale=0.38]{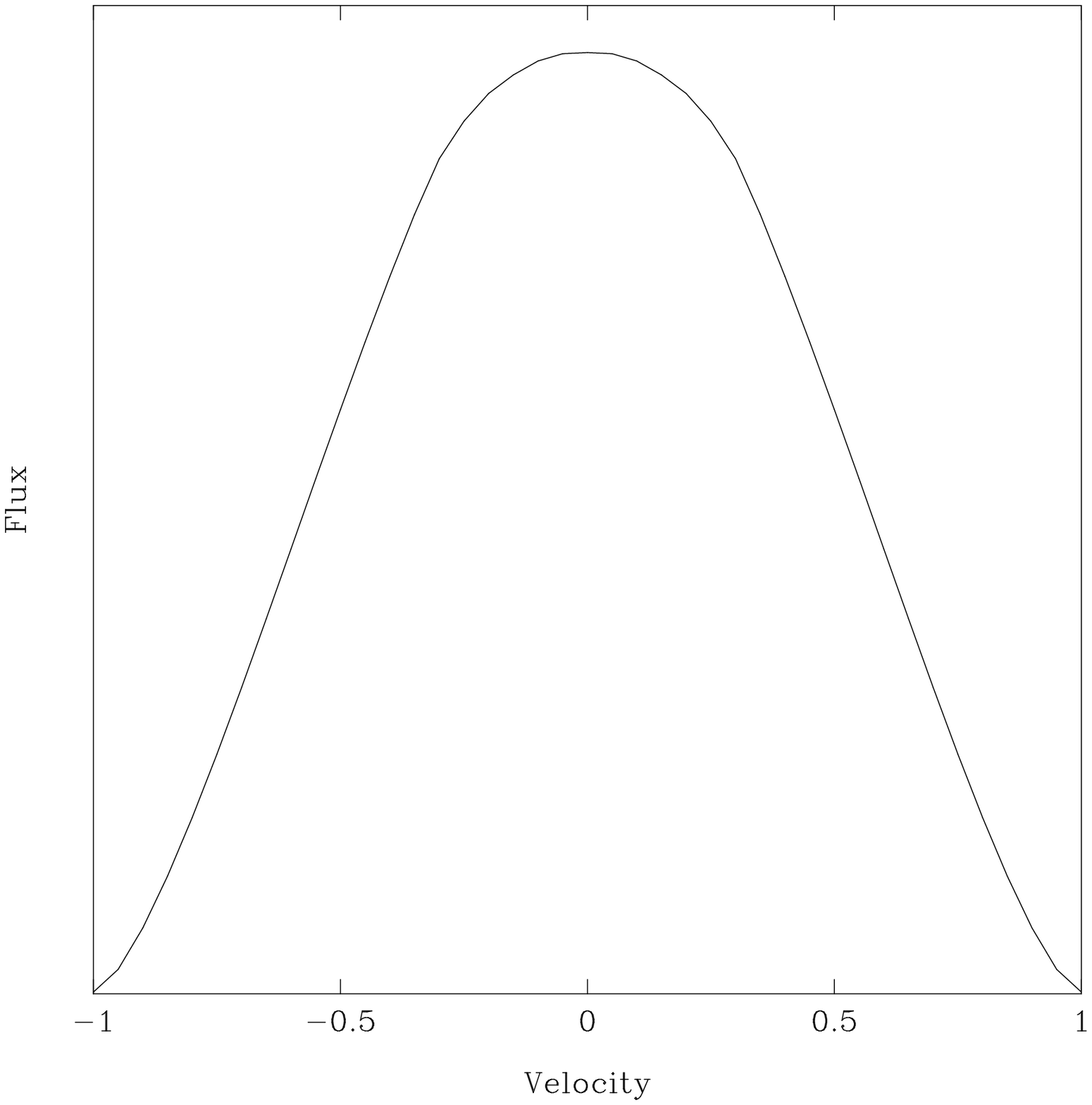}\label{bicone spectrum}}\\
\subfigure[Hollow bicones,  inclined 35$\,^\circ$ RA$=28\,^\circ$]{\includegraphics[scale=0.38]{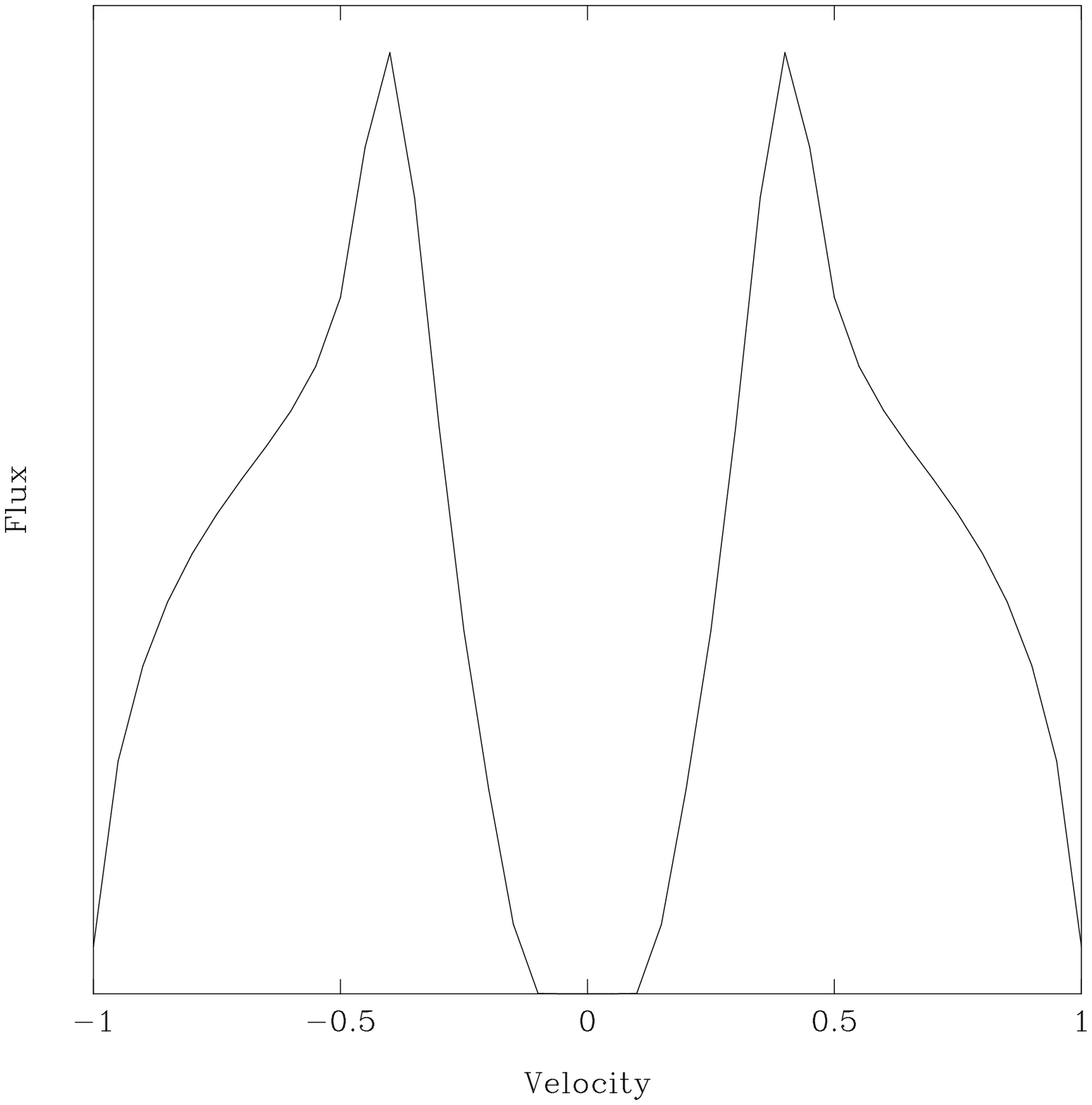}\label{hollow bicone spectrum}}
\subfigure[Disk, inclined 45$\,^\circ$, RA$=45\,^\circ$]{\includegraphics[scale=0.38]{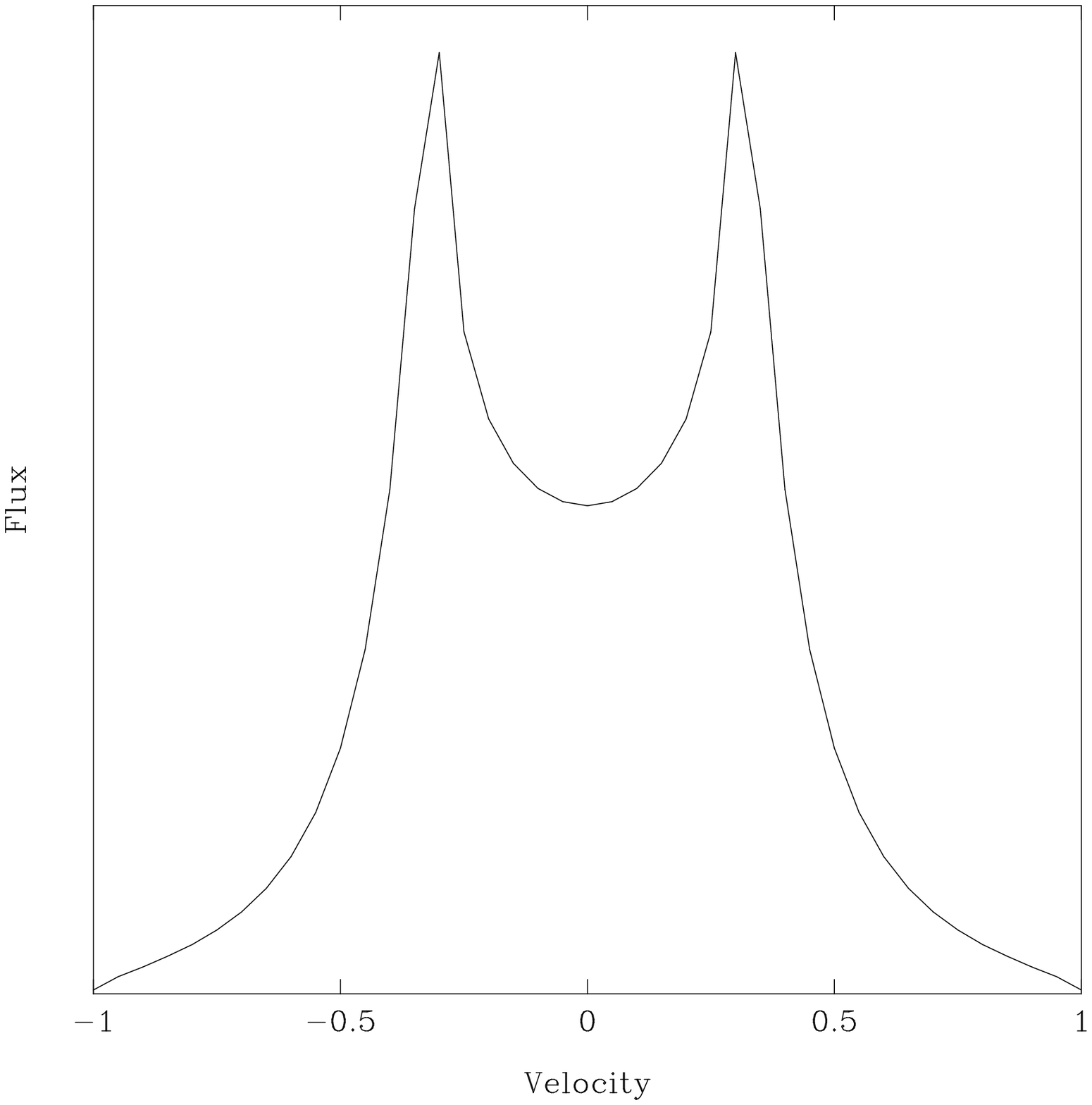}\label{disk spectrum}}
\caption{The spectra for the oriented source models shown in Figure \ref{models}. Velocities are normalized to -1 to 1
for each model regardless of orientation. The flux is a normalized value which is not shown, only the spectrum shape is 
significant.}
\label{spectra}
\end{figure*}

\subsubsection{Sphere}

The sphere is the simplest model, and is depicted in Figure \ref{sphere profile}.
The material is inflowing, with $p = 0.5$ in Equation \ref{eqn:velocity}, so that it is decelerating.

\begin{figure}
\centering
\includegraphics[scale=0.48]{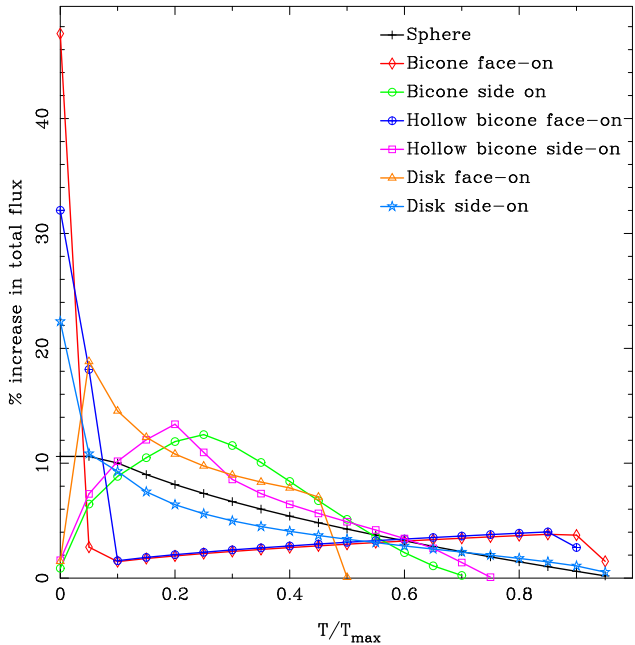}
\caption{Change in observed BELR brightness over time in a flaring BELR, for the  models at several orientations, with the flaring  material having twice the emission than it does in its quiescent state.  The vertical axis is the percentage increase in flux of the total  BELR due to the flare, the horizontal axis is the time that has passed. Time has been binned into  intervals that are 5\% of the flare lifetime. Side-on orientations are inclined by 90$\,^\circ$.}
\label{flare percent}
\end{figure}
 
\subsubsection{Bicones}
Two mirror-image cones are cut out from the sphere to form a biconical structure. The opening angle of the cones is 30$^\circ$. The cones may be either solid, or hollow, in the latter case consisting of  a 
conical shell where the shell wall has an opening angle of 3$^\circ$. The direction of velocity is reversed so that the inflow becomes an accelerating outflow.
The hollow cone  is intended to be combined with the disk to form a disk-wind model, in which case it should have have a rotational velocity component, but that is not used here. The cones may be inclined and then rotated on the plane of the map, with the anchor point  the midpoint between the bicones. 
An inclination of 0$^\circ$ means that one of the cones is pointing directly towards the observer, and the other is pointing away. 
The rotation of the model is specified as a rotation angle (RA), where $0\,^\circ$ rotation is pointing down the page  on the magnification map in Figure \ref{map}, and $90\,^\circ$ is pointing to the right.  Figure \ref{bicone profile}  depicts the solid cones, inclined by 90$^\circ$, RA$=0\,^\circ$. Figure \ref{hollow bicone profile} is slightly more complex, it shows the hollow cones, inclined by  35$^\circ$, RA$=45\,^\circ$. 

\subsubsection{Keplerian Disk}

This model is a flat rotating disk with the BELR material in Keplerian orbits, so $p = -0.5$ in Equation \ref{eqn:velocity}.  The thickness of the disk is 1\% of the outer radius.  Like the bicones, the disk can be oriented with respect to the observer. By convention, an inclination of 0$^\circ$ is face-on. Figure \ref{disk profile} shows the disk inclined by 45$^\circ$, RA$=45\,^\circ$.

\subsection{Flaring}
Within a certain time interval some material will be observed to be flaring. That material can be found based on the temporal properties of the flare discussed in Section \ref{sec:reverb}. 
The  emissivity of the   material is increased from $\epsilon$ (Equation \ref{eqn:emiss}) to $k\epsilon$. 
A suitable value for $k$ must be chosen;   \citet{kaspi+07} have observed increases of 10-70\%
in the total BELR flux in several quasars at some (unknown) time during a flare. 
If we numerically determine the flux over time for a flare in all the models, at some different orientations,  when $k = 2$,  the result is shown in Figure \ref{flare percent}.
The horizontal axis is the percentage of the lifetime of the flare, the vertical axis is the percentage increase in flux of the BELR. 

This shows that the most increase in brightness occurs in the early stages of the flare,  due to the time delay surface moving away from the observer as depicted in Figure \ref{constant_vlos}. The face-on cones show the biggest increase  because the material along the line-of-sight is observed to flare first and this is a substantial amount of the total material in the face-on cones. The side-on orientations brighten  later as they have less material along the line-of-sight. To estimate a flux increase for our models, we use the sphere, which has a increase of 10\%, roughly  the mean of all the models. To be consistent with the observations of \citet{kaspi+07}, we conservatively set $k = 5$ for all models, which produces an increase of 50\% in the sphere in the early stages.

\subsection{Model slicing}

Once a model has been generated with the appropriate velocity and emissivity profile, it is sliced by time to produce many profiles representing the progress of the flare.  To find how microlensing  changes the spectrum, each time slice must be further sliced by velocity. The resulting slice profiles are microlensed and the unlensed and lensed flux of the slice recorded.  This produces a grid of values indexed by lensed/unlensed, time, and velocity. Spectra for flares can be implemented  by selecting and combining values from the grid. For example, the lensed spectrum of the flaring BELR at time T is obtained by multiplying the lensed velocity slices for time T by $k$, and combining that with all other lensed grid values into a total flaring BELR spectrum.

Spectra for the quiescent BELR of Figure \ref{models} are shown in Figure \ref{spectra}. These spectra are consistent with those reported in other works \citep[e.g.][]{abajas+02}, but not identical, because of the different radii, thicknesses, and other values used here.

We use 12 time intervals for the purpose of visualization. The flare is brightest in the early times, and also changes rapidly then, so the first two  time intervals are shorter, i.e. of higher time-resolution, than those that follow. The flare also has a small profile at the end of its life, and therefore possibly high microlensing variability, so the last two time intervals  will be shorter than those preceding them. For a BELR of size 0.06 pc in Q2237+0305, a flare will have a duration of $142.95~h_{70}^{-1/2}(M/\rmn{M}_\odot)$~days in the rest frame of the quasar.
 The lensed quasar in Q2237+0305 is located at a redshift of $z_S = 1.695$, so the flare will have a duration of $385.26~h_{70}^{-1/2}(M/\rmn{M}_\odot)$~days in the observer's frame. Therefore we set $T_{max} = 385$ days as the lifetime  of a flare through our BELR models; the time intervals begin at 0, 0.05, 0.1, 0.2  ... 0.8, 0.9,  0.95 of $T_{max}$. All the material that is flaring within the time interval will be used to form a brightness profile for that interval.
For the spectra we use 40 velocity slices within each time slice, a value chosen to give good velocity resolution for data analysis and visualization. The velocities are
normalized to be between -1 and 1 regardless of the profile's orientation.

\subsection{Model orientations and map location}

\begin{figure*}
\centering
\subfigure[format=hang][Sphere]{\centering\includegraphics[width=41mm]{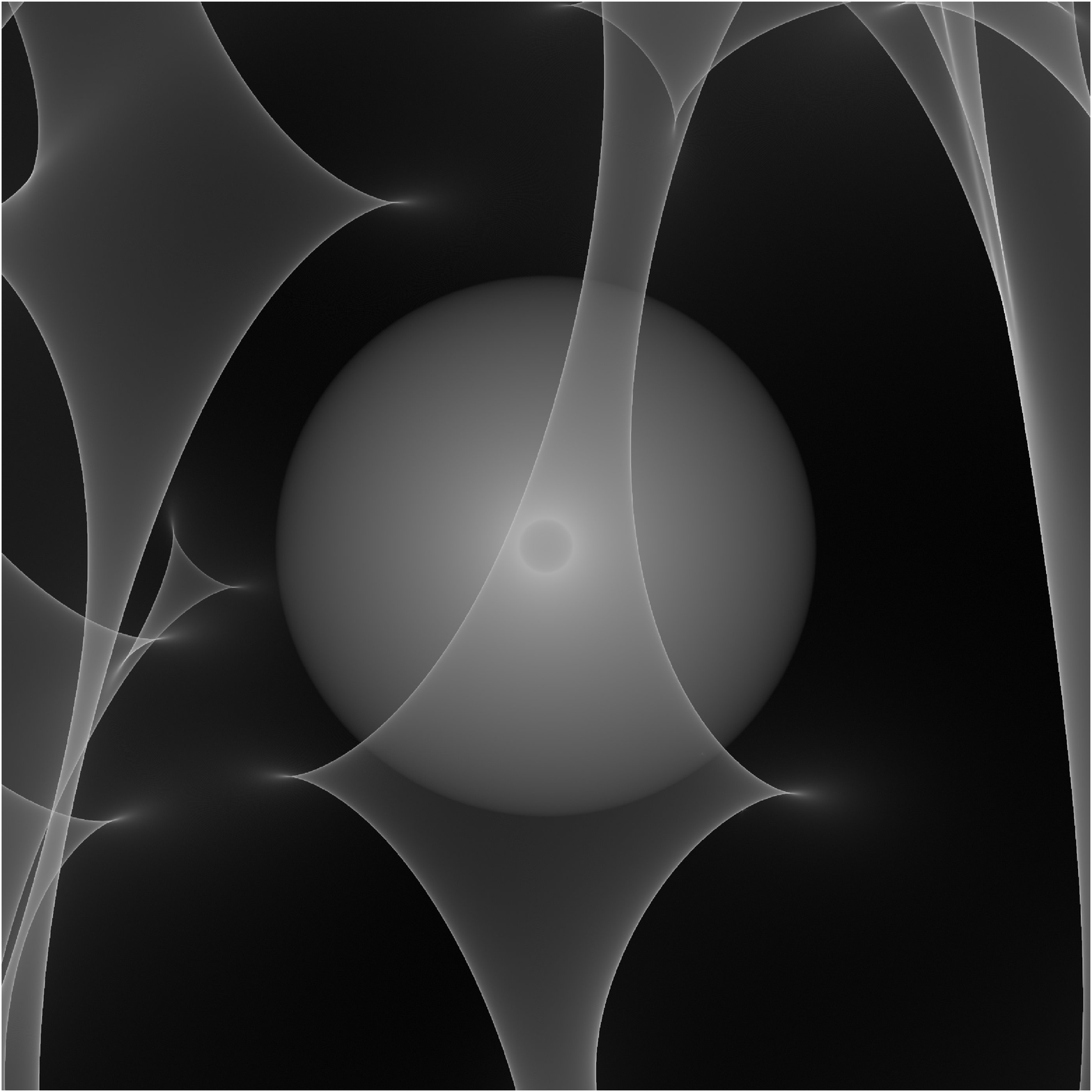}\label{sphere in situ}}
\subfigure[format=hang][Solid bicones, face--on]{\centering\includegraphics[width=41mm]{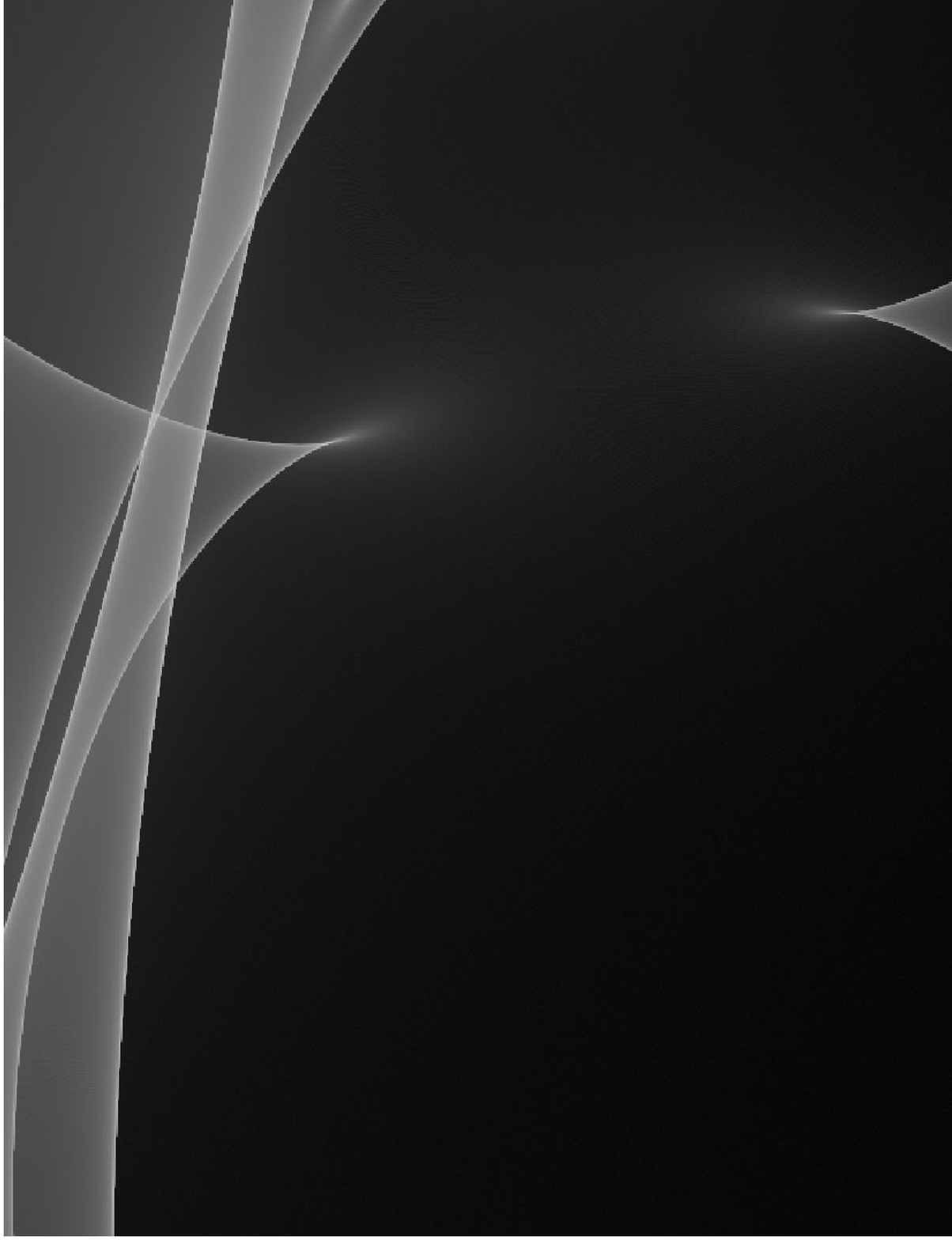}\label{bicone 0 in situ}}
\subfigure[format=hang][Solid bicones, inclined  $60^\circ$, RA$=30\,^\circ$]{\centering\includegraphics[width=41mm]{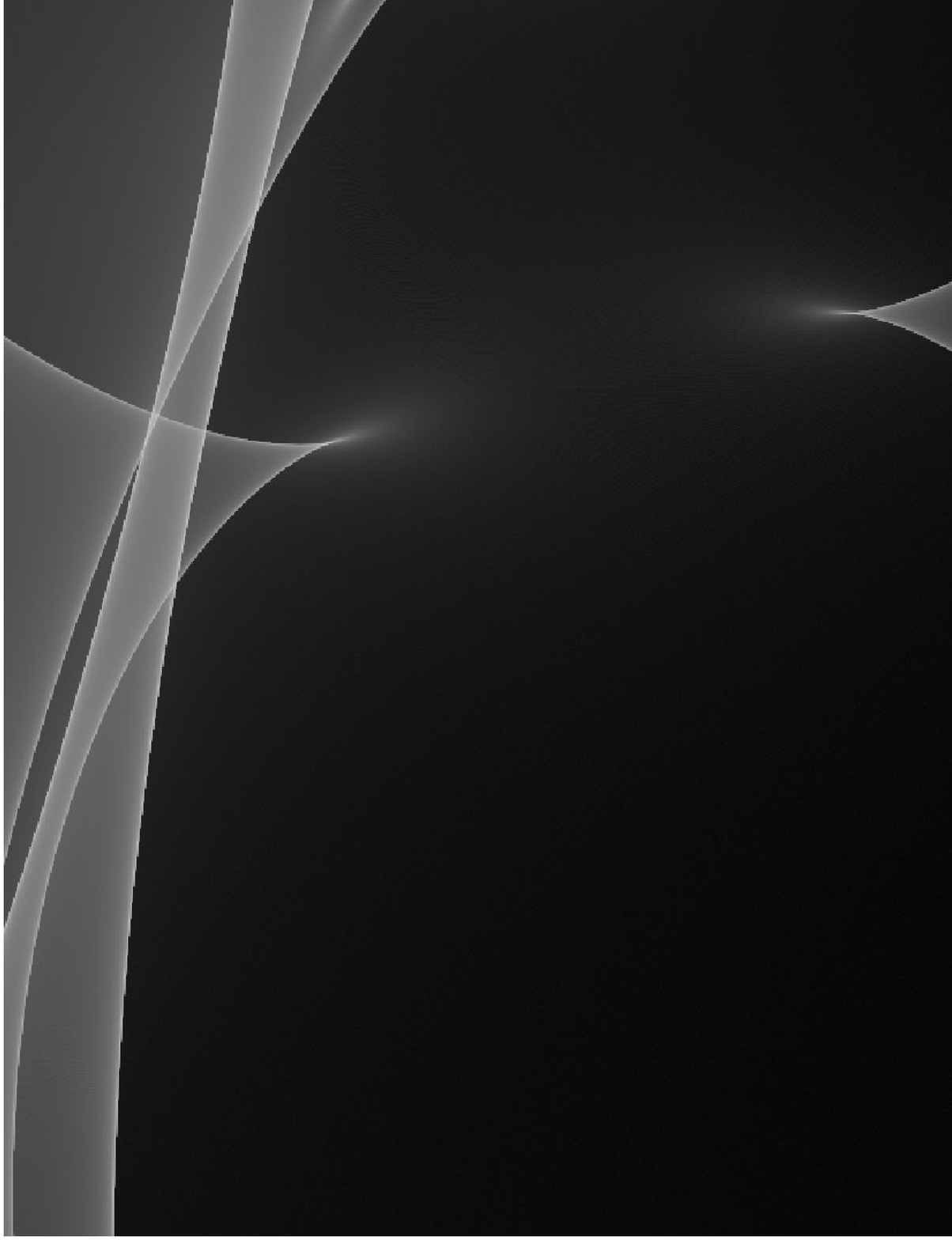}\label{bicone 60 in situ}}
\subfigure[format=hang][Hollow bicones, face-on]{\centering\includegraphics[width=41mm]{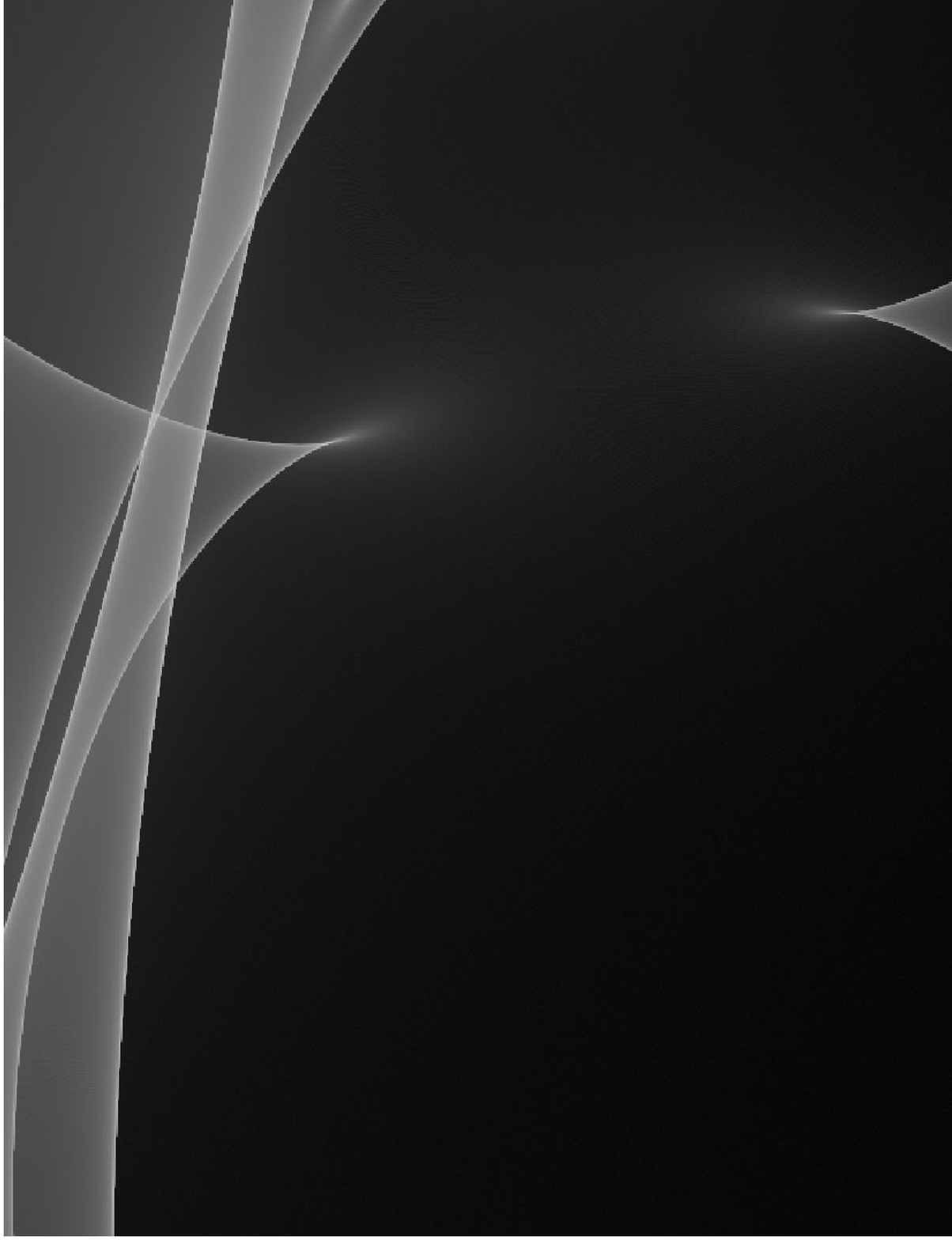}\label{hollow bicone 0 in situ}}\\
\subfigure[format=hang][Hollow bicones, inclined  $60^\circ$, RA$=30\,^\circ$]{\centering\includegraphics[width=41mm]{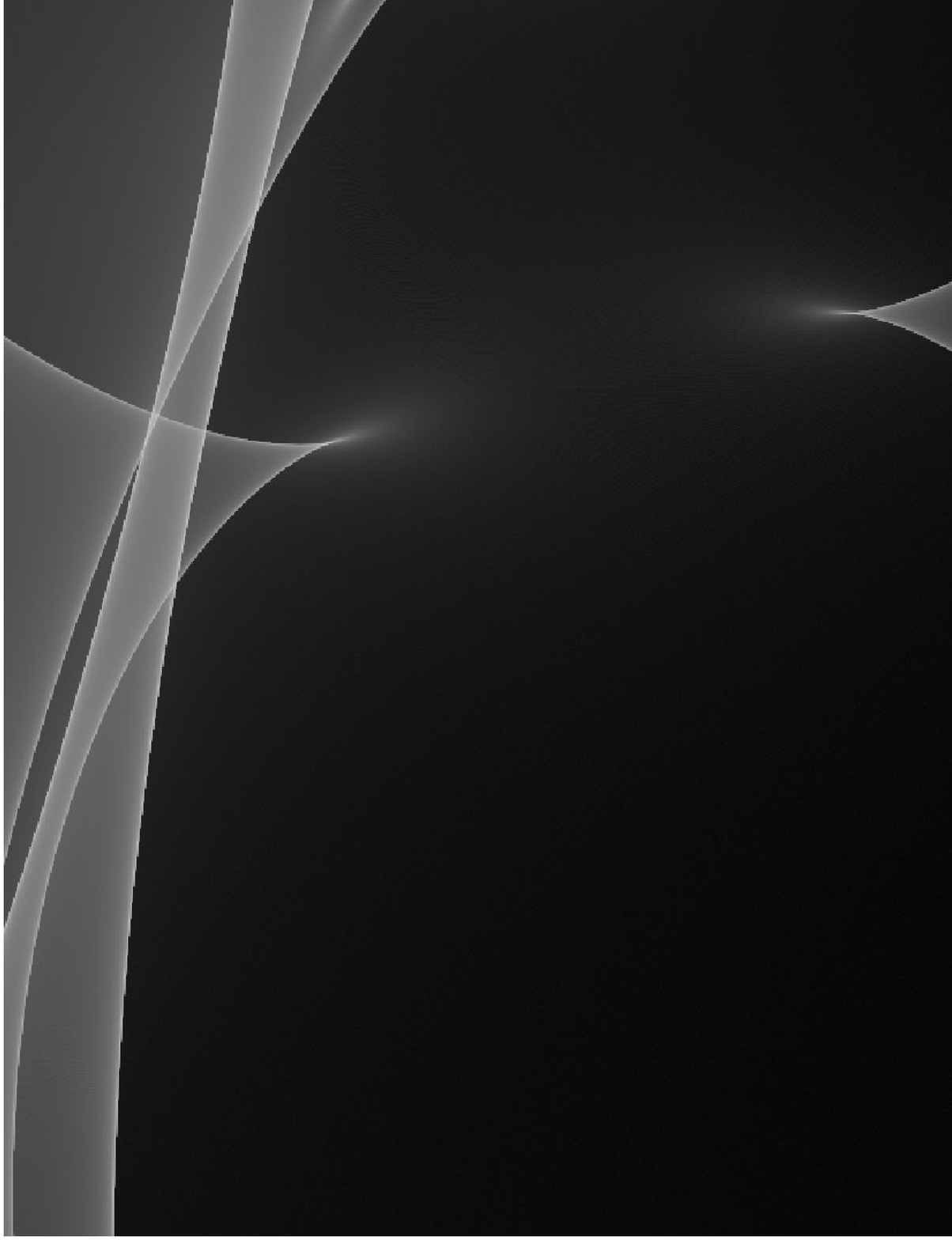}\label{hollow bicone 60 in situ}}
\subfigure[format=hang][Disk, inclined $10^\circ$, RA$=30\,^\circ$]{\centering\includegraphics[width=41mm]{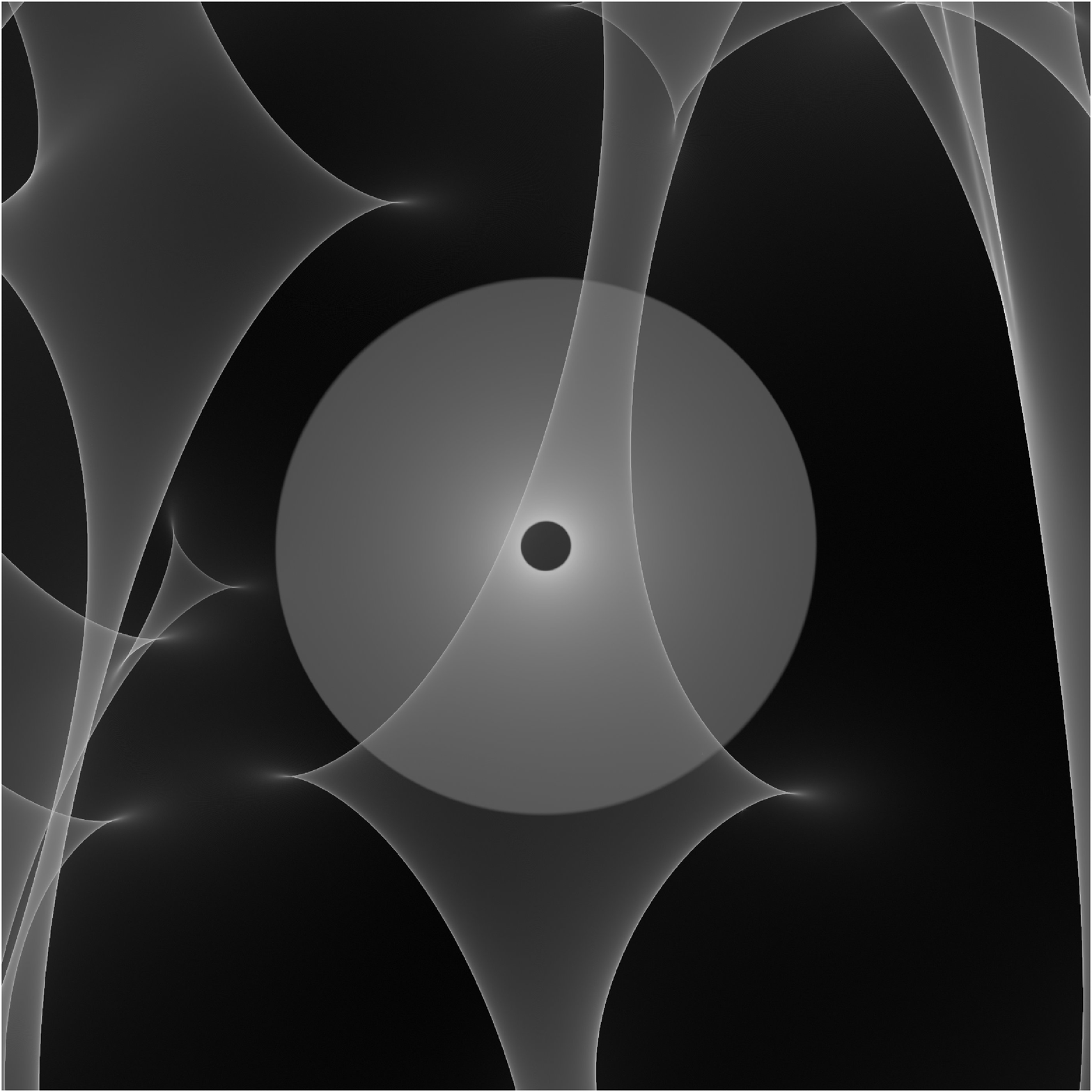}\label{disk 10 in situ}}
\subfigure[format=hang][Disk, inclined $60^\circ$, RA$=30\,^\circ$]{\centering\includegraphics[width=41mm]{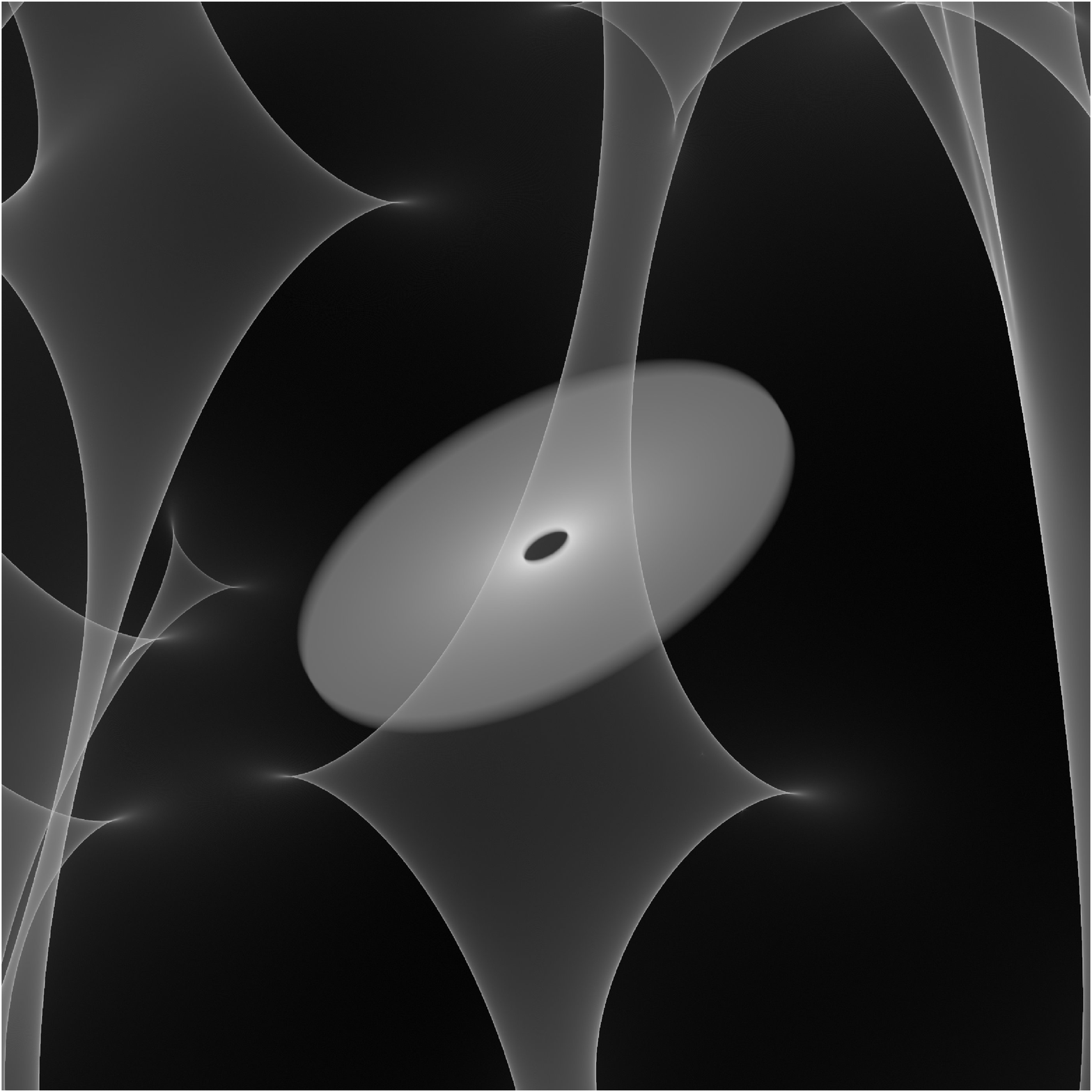}\label{disk 60 in situ}}
\caption{Source profiles superimposed on a location of the map shown in Figure \ref{map}. These are the profiles and
magnification map  location  used to generate the data in this paper. Note that the orientations are not the same as Figure \ref{models}. The models have a  radius of 0.06 pc (1 ER); the section of the map that is shown here has a width of 0.24 pc (4 ER). The disk is rotating clockwise so that the bottom-left is towards the observer (blue-shifted, negative velocities).}
\label{locations}
\end{figure*}

 In this paper we will not attempt to present statistical  results for many orientations and locations, but focus on the behaviour for a simple situation, thus demonstrating the method and indicating possible fruitful directions of investigation. Figure \ref{locations} shows the source models with the orientations we have chosen, superimposed on the magnification map at the  location  we have chosen. The sources lie  over a region between two caustics where the magnification is increased; note that the material close to the center of the BELR, which has the highest emission,   will be magnified more than the outlying  material. The caustics will be referred to as the left and right caustics, i.e. as they appear on the map.  Some of the profiles extend to the bottom right where the right caustic proceeds towards a cusp. The orientations that we use are either wholly face-on (0$\,^\circ$ inclination) or mostly side-on: $60^\circ$ inclination. The disk when face-on (Figure \ref{locations} (f)) has no line-of-sight velocity, therefore it is at $10^\circ$ inclination so it has some line-of-sight velocity components.

\section{Qualitative results}
\label{sec:quals}

\begin{figure*}
\includegraphics[width=500pt]{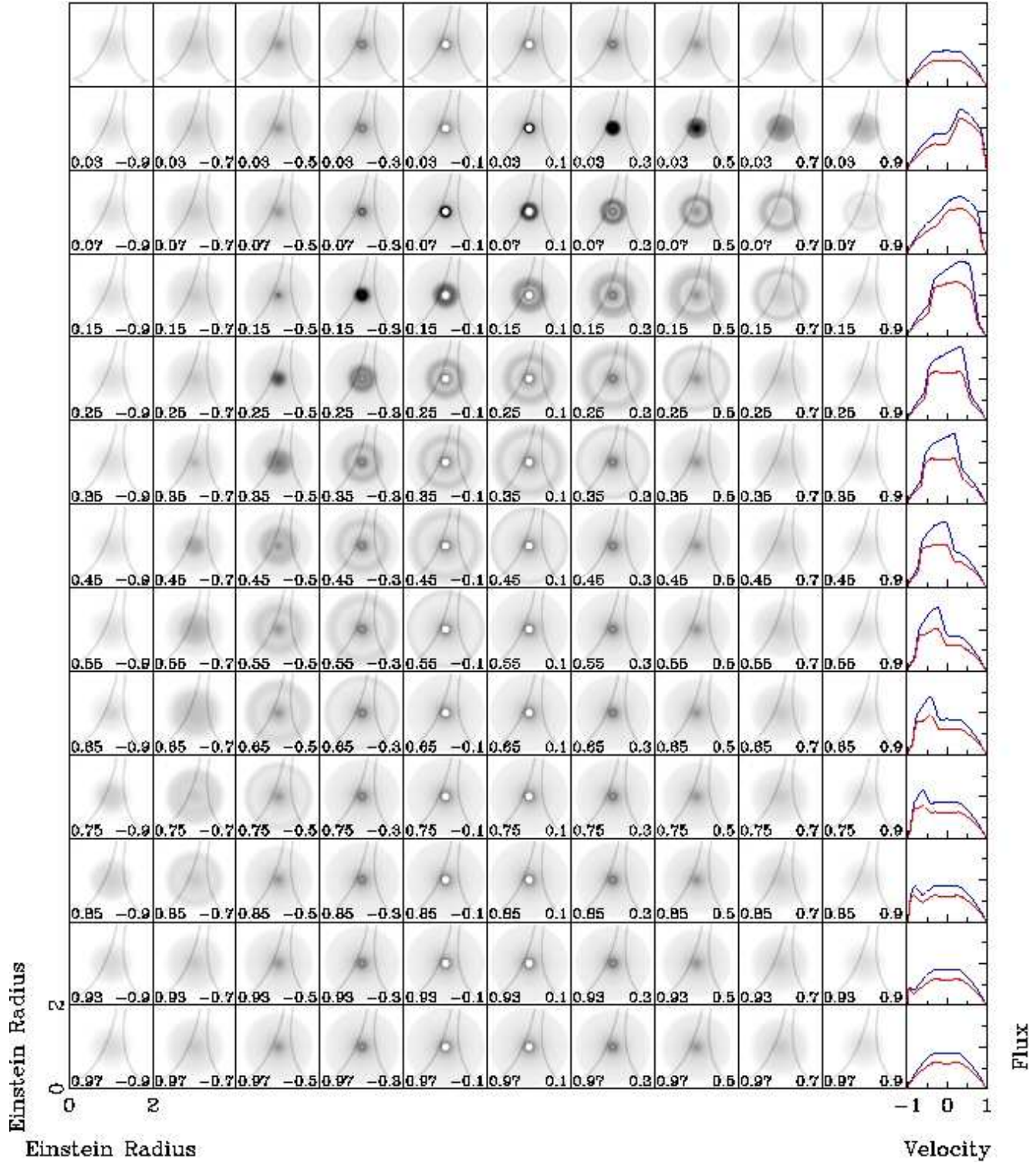}
\caption{Images of the surface brightness profiles (first 10 columns) and spectra (last column) of the spherical inflowing BELR (Figure \ref{locations}(a)) as a flare propagates through it. The profiles are superimposed over the outline of the caustic structure from the magnification map. The images are time and velocity sliced profiles of the  BELR; the first number in a BELR image is the central time, normalized to $T_{max}$, of the time interval, the second number is the central velocity of the velocity interval.  The first 10 columns contain the BELR profile velocities in slices of 0.2,  beginning at   -1, -0.8, -.6, ... 0.8,  1, from left to right. The first row shows the quiescent BELR. Subsequent rows, reading down the page, show the flaring BELR  sliced into time intervals beginning at 0,  0.05,  0.1, 0.2,  0.3 ...
 0.8, 0.9,  0.95 of $T_{max}$. The flaring material  appears darker than the rest of the BELR; note that it has been rescaled for enhanced visualization in the profiles. The column on the right shows the spectrum of the BELR, the blue  line is the unlensed spectrum, and the red line is the lensed spectrum. The lensed spectra have been scaled relative to the mean magnification, for comparison with the unlensed. Each spectrum in a row is the spectrum of the flaring BELR for the time interval of the row. Note how the flare appears at high positive velocities (right), and over time (down the page) migrates  to negative velocities; this is also tracked in the spectra. }
\label{sphere strips}
\end{figure*}

\begin{figure*}
\includegraphics[width=500pt]{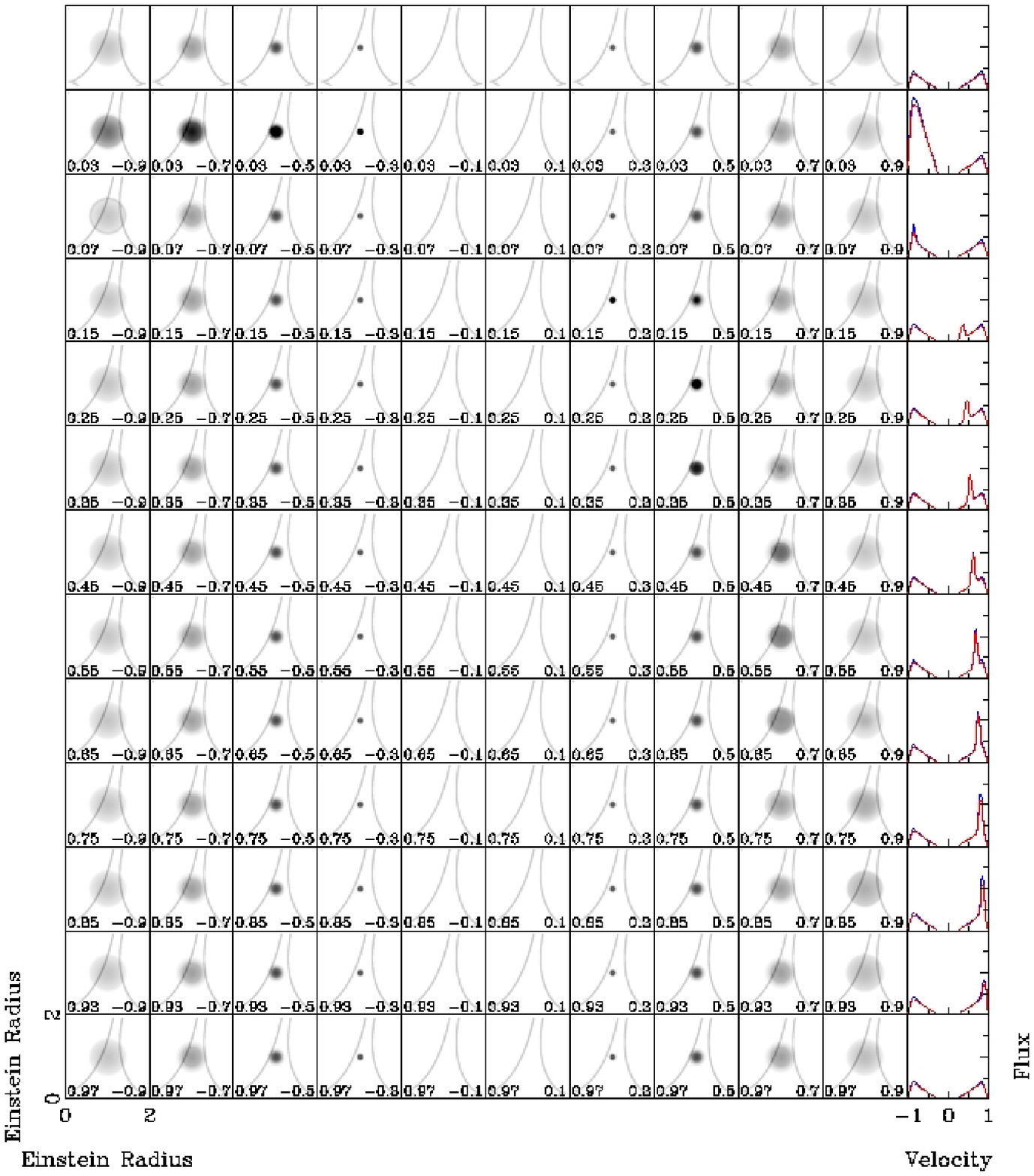}

\caption{As for Figure \ref{sphere strips}, but the source model is the face-on solid bicones (Figure \ref{bicone 0 in situ}).}\label{bicone 0 strips}
\end{figure*}

\begin{figure*}
\includegraphics[width=500pt]{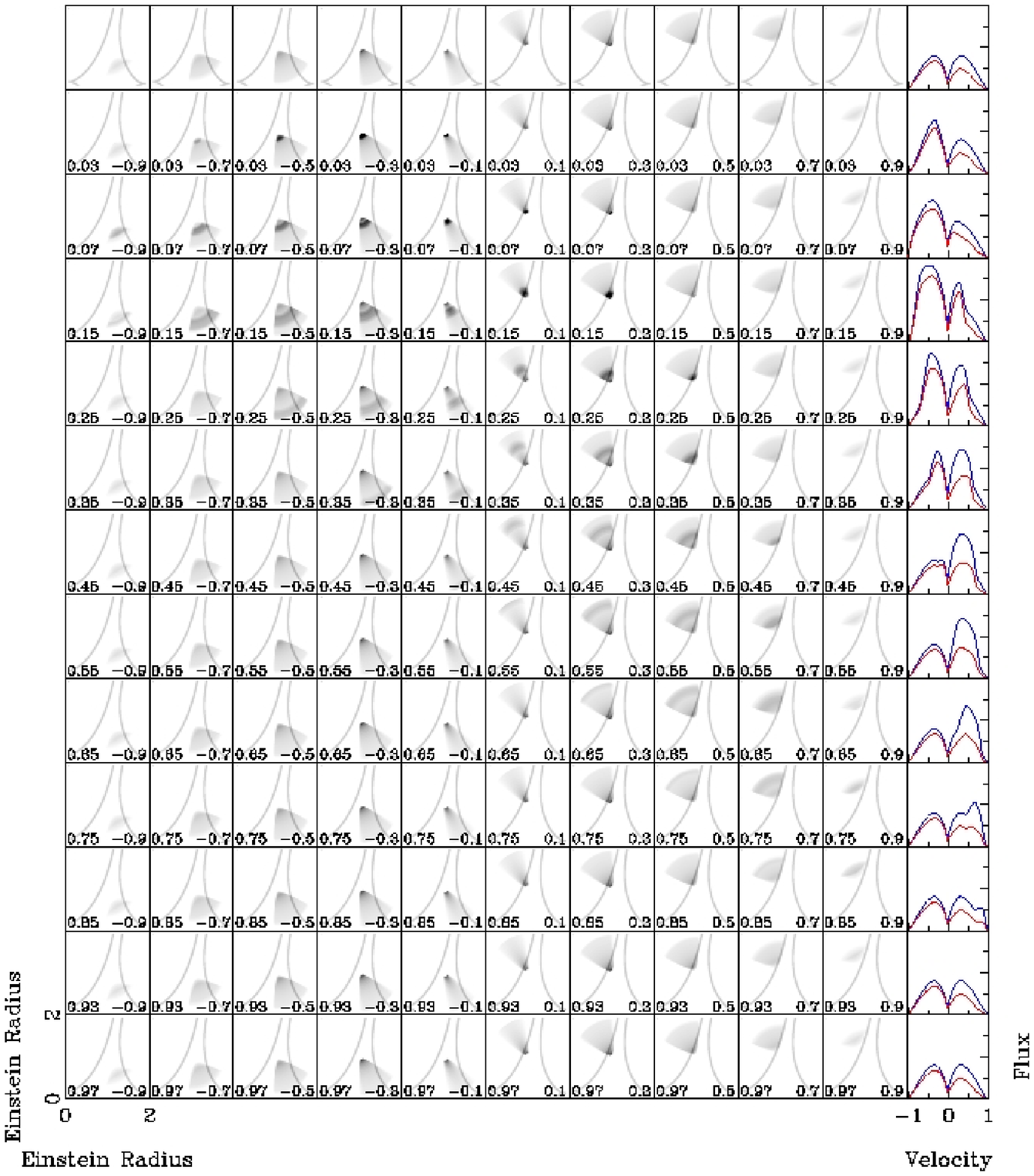}

\caption{As for Figure \ref{sphere strips}, but the source model is the side-on solid bicones (Figure \ref{bicone 60 in situ}).}\label{bicone 60 strips}
\end{figure*}

\begin{figure*}
\includegraphics[width=500pt]{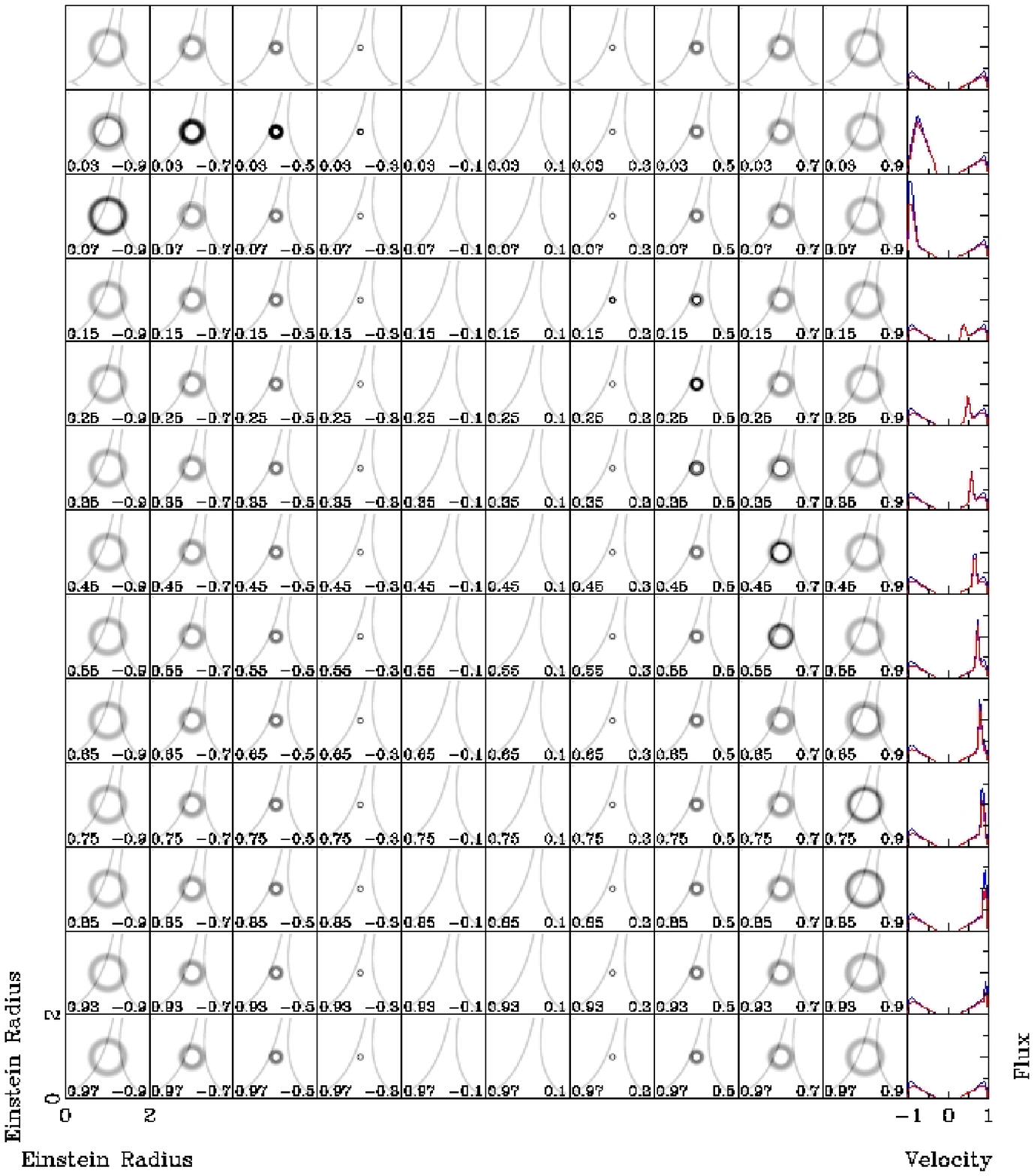}

\caption{As for Figure \ref{sphere strips}, but the source model is the face-on hollow bicones (Figure \ref{hollow bicone 0 in situ}).}\label{hollow bicone 0 strips}
\end{figure*}

\begin{figure*}
\includegraphics[width=500pt]{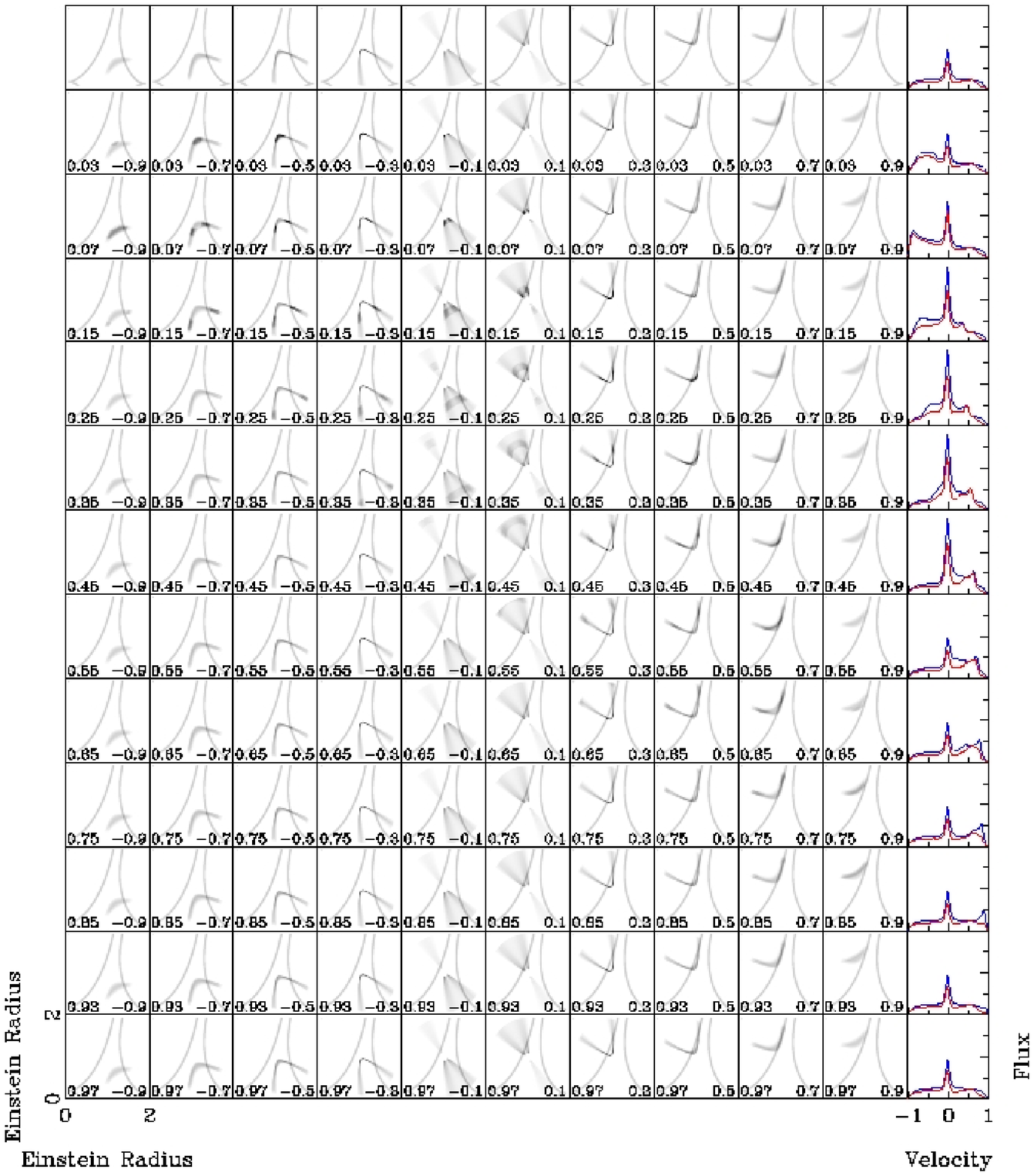}

\caption{As for Figure \ref{sphere strips}, but the source model is the side-on hollow bicones (Figure \ref{hollow bicone 60 in situ}).}\label{hollow bicone 60 strips}
\end{figure*}

\begin{figure*}
\includegraphics[width=500pt]{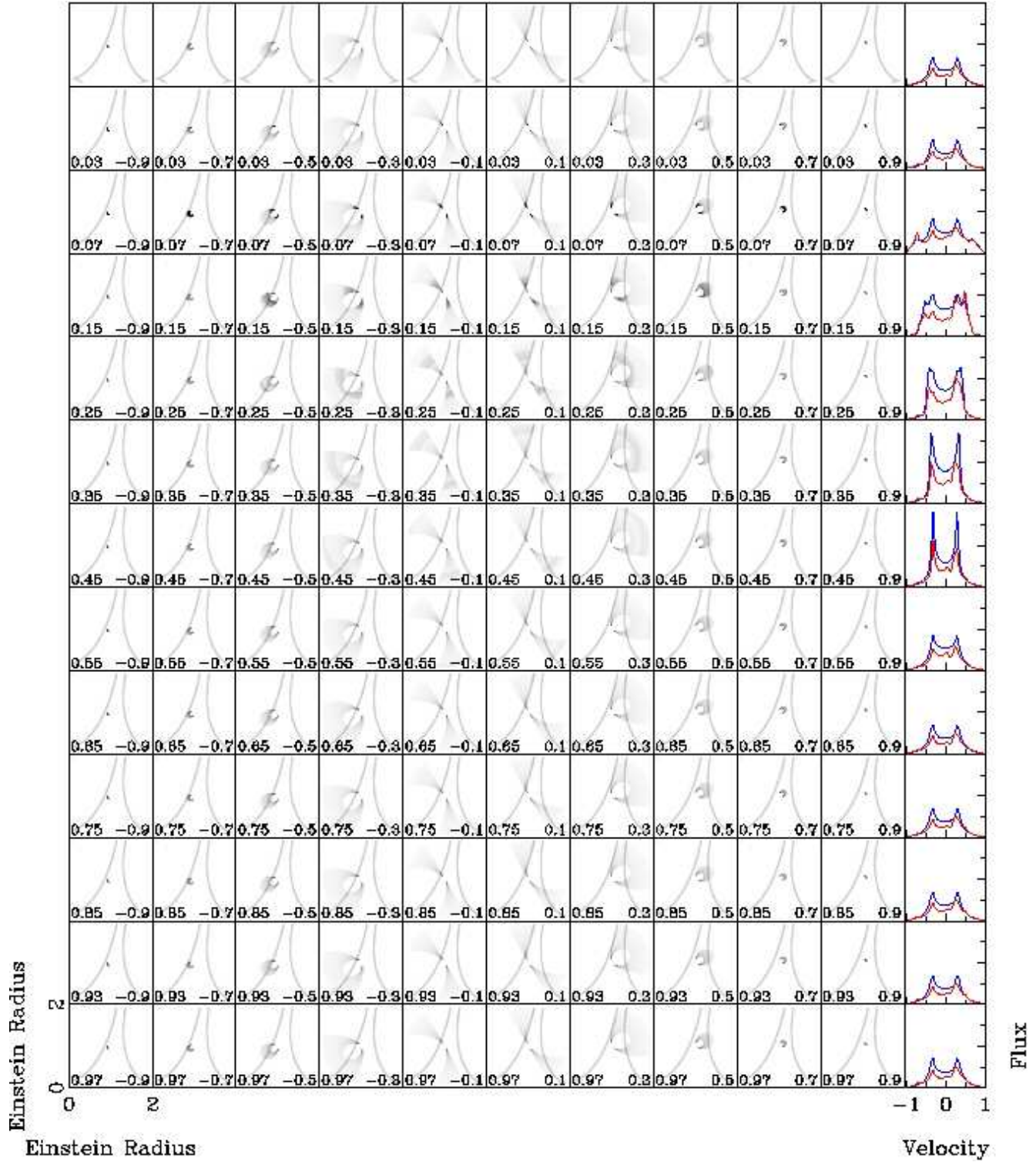}

\caption{As for Figure \ref{sphere strips}, but the source model is the almost-face-on disk(Figure \ref{disk 10 in situ}).}
\label{disk 10 strips}
\end{figure*}

\begin{figure*}
\includegraphics[width=500pt]{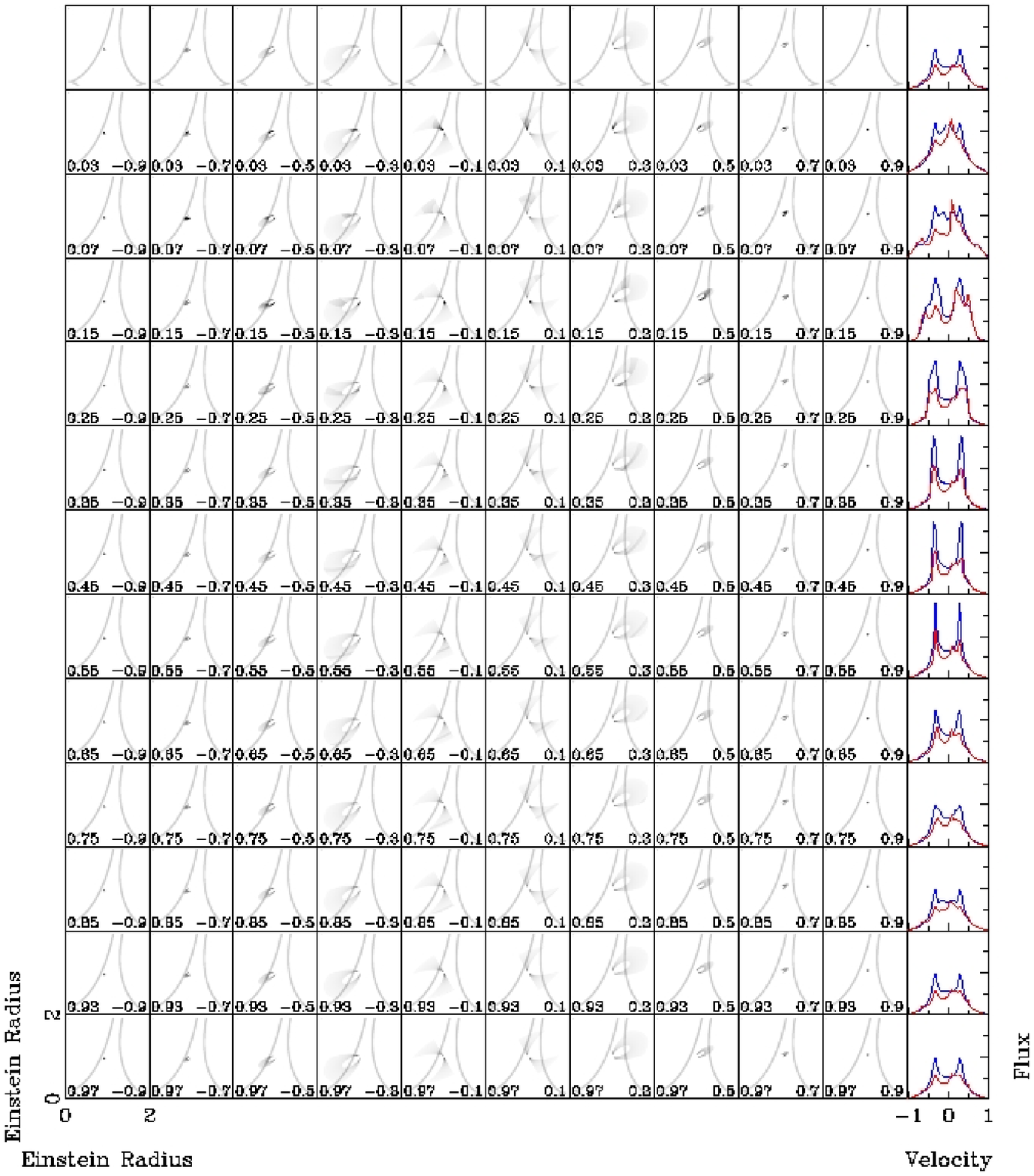}

\caption{As for Figure \ref{sphere strips}, but the source model is the side-on disk (Figure \ref{disk 60 in situ}).}
\label{disk 60 strips}
\end{figure*}

We present source profiles and spectra of a BELR as a flare progresses through it.
Figure \ref{sphere strips} shows the source brightness profile of a flare propagating through the spherical inflow BELR (Figure \ref{models}(a)). The profiles are superimposed over the outline of the caustic structure from the magnification map. Each column shows the BELR profile sliced by velocity, each row shows the profile within a time interval. The velocity bin for each column has a width of 0.2, beginning at  -1, -0.8, -0.6, ... 0.8, from left to right column. The first  row shows the BELR before the flare begins, i.e. the quiescent BELR. The time intervals for the subsequent rows, reading down the Figure, begin at 0,  0.05, 0.1, 0.2 ... 0.9, 0.95 of $T_{max}$, the lifetime of the flare. The image intensity indicates the amount of emission, so that the flare is visible as dark patches, note the flare has been rescaled in the images for ease of visualization.  The two numbers in each box show the central velocity and time of the BELR material in the image. The right-most column is the unlensed and lensed spectrum of the flaring BELR in the row, constructed from 40 velocity bins, i.e. a higher resolution than can be shown in the source profiles. The spectrum in  blue  is the unlensed spectrum, the lensed spectrum is in red. The lensed spectrum has been scaled to the mean magnification of image A, for comparison with the unlensed. The spectrum velocities,  indicated on the horizontal axis of the bottom spectra, are normalised values from -1 to 1. The flux values (vertical axis)  are  arbitrary and are not indicated.
Figures \ref{bicone 0 strips} -- \ref{disk 60 strips} show the flaring BELR for the other source models.


\section{Discussion}
\label{sec:discuss}

Of  interest is the comparison of the unlensed and lensed spectrum as the flare propagates, and whether that can be used to differentiate between  BELR models, and their orientations. Here we will present a description of the spectra for the various source models at the location and orientations we have chosen. By using a single location with simple caustic structure we demonstrate that microlensing can alter the  spectrum of a flare in the BELR, and the sorts of signatures we might expect to see. Future contributions will provide a comprehensive analysis describing the microlensing behaviour  for many locations on the magnification map.

\subsection{Comments on the microlensing of the quiescent BELR profiles}

\begin{table}

 \caption{Total magnification of the sources after microlensing, at the location chosen (Figure \ref{locations}).}

 \begin{tabular}{ccccccc}
 \hline
  Profile &  Magnification \\

   \hline
   Sphere & 0.75 \\
   Bicones inclined 0$\,^\circ$ & 0.93 \\
   Bicones inclined 60$\,^\circ$ & 0.70 \\
   Hollow bicones inclined 0$\,^\circ$ &  0.86 \\
   Hollow bicones inclined 60$\,^\circ$ & 0.74 \\
   Disk inclined 10$\,^\circ$ & 0.72 \\
   Disk inclined 60$\,^\circ$ & 0.76 \\
  \hline
 \end{tabular}

\label{magnifications}
\end{table}

At the location chosen, the source is demagnified relative to the mean magnification; Table \ref{magnifications} lists the magnifications for the various profiles, scaled to the mean  of image A. The magnifications are around 0.75, except for the face on cones, which are higher because most of the flux for these profiles sits within the high magnification region (Figure \ref{locations}). The spectra for the microlensing of the
profiles are in the top row, last column, of Figures \ref{sphere strips} -- \ref{disk 60 strips}. In all cases the lensed spectrum  retains a similar shape to the unlensed spectrum. The inclined solid bicones (Figure \ref{bicone 60 strips}) have more material towards the ends, which counters the decline in  emissivity, and the spectrum has a rounded peak. The spike at 0 velocity for the inclined hollow cones (Figure \ref{hollow bicone 60 strips}) is due to their orientation: with an opening angle of 
30$\,^\circ$ and an inclination of 60$\,^\circ$, each cone has an edge that is lying flat along the plane of the sky, the material there has 0 line-of-sight velocity.


\subsection{Comments on the effect of the flare}
Now we turn to the other rows in Figures \ref{sphere strips} -- \ref{disk 60 strips}, those that display the behaviour of the flare. Where necessary we will refer to the row by its central time, eg. T=0.07, and column by central velocity, eg. V=0.5. Some of the spectra show effects caused by the microlensing of the flare, we will highlight and discuss the interesting ones.

The sphere and the face-on bicones do not show much difference between the unlensed and lensed spectrum over time, because of the symmetries in these models. Blue- and redshifted material is spatially co-located at each point of the profile, and also at a range of velocities between the extremes of -1 to 1. This means there will be little differential microlensing. 
A small effect in the sphere (Figure \ref{sphere strips}) is the  enhanced sharpness of the peak at T=0.75 in the lensed spectrum  at V=-0.8, which occurs because the flare is small at this point and passes over the left caustic, near the top left  of the  sphere inner radius. There are some effects with the hollow bicones (Figure \ref{hollow bicone 0 strips}), due to the cone edge lying over a caustic.  As the flare moves down the edges of the face-on cone it produces a peak at negative velocities, at T=0.03-0.07,
which is diminished in the lensed spectra because some material lies outside the left caustic. As the flare moves off the cone edge and inside the caustics,  the peak disappears in both spectra and they become similar at the negative end (T=0.35).
Corresponding behaviour can be seen in positive velocities from  T=0.15 as the   flare travels along the edge of the away-facing cone.

The inclined bicones produce more interesting effects due the flare passing over caustics. 
The ``kink'' at  T=0.15, V=0.5 in the lensed spectrum for the inclined solid bicones (Figure \ref{bicone 60 strips}) is not a drop at high velocities but a raising of  lower velocities, because the flare is sitting on the left caustic near the base of the red-shifted cone,  and low positive velocities are magnified.
Similarly, the ``dimple'' in  the peak at positive velocities for T=0.75-0.85 is  really a second peak around V=0.6. At this time interval the surface of the flare cuts down and partly away from the observer,  through the redshifted cone. The distant side of the cone (as viewed by the observer) has faster positive velocities, and some of this material is near the left caustic, and therefore magnified.  As with the solid bicones, the flare in the hollow bicones (Figure \ref{hollow bicone 60 strips})  lifts both the lensed and unlensed spectra at the negative end, but by T=0.25 the lensed flux has dropped. At T=0.15, V=-0.5 the source profile consists of  two edges and the flare is on both of them,
but one  is outside the high magnification region, leading to the drop in flux. A similar effect occurs at positive velocities where only one of the two edges of the away-facing cone is close to the left caustic.

\begin{figure}
\includegraphics[scale=.235]{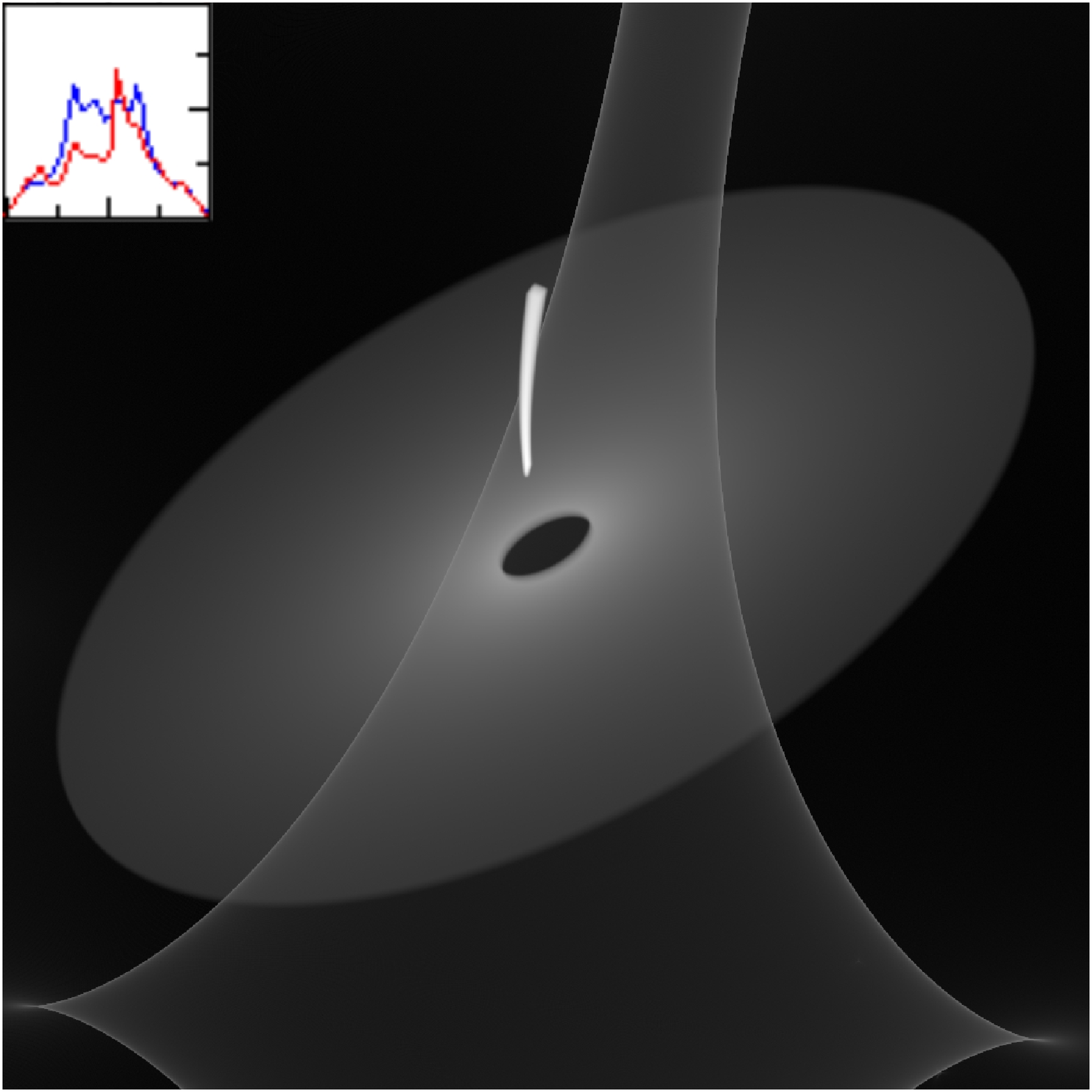}  
\caption{Showing the flaring material in the inclined disk (Figure \ref{locations}(g)) at T=0.05-0.1 and velocity  $0.08 \le V \le 0.09$, superimposed on the map. The region of the map shown  has width 0.12 pc  (2 ER). The flare material is brightened relative to the rest of the disk, for visualization, and is shown in white. The spectrum from row 
T=0.07, Figure \ref{disk 60 strips}, in which there is a high magnification  spike produced by this material, is shown at the top left.
The material is highly magnified because it sits on a caustic. }
\label{disk part}
\end{figure}

Both face-on and inclined disk profiles produce effects.
In the face-on disk spectra (Figure \ref{disk 10 strips}) at T=0.07, the flare is in  small regions lying on both caustics, so  multiple small peaks are produced in the lensed spectrum. By T=0.25 these peaks have disappeared and at negative velocities the  lensed spectrum is diminished because the blue-shifted half of the disk, particularly that close to the core, is less magnified.  
Of interest in the inclined disk (Figure \ref{disk 60 strips}), and the most significant effect presented here, is the large spike at small positive velocities at T=0.03-0.15. This is due to the flare reaching some disk material at points just after it has turned over from negative to positive velocities, where it is sitting on a caustic, and highly magnified. This material can be seen
in Figure \ref{disk part} inside the inclined disk, and is brightened in the disk for visualization purposes. The material lies within the time interval T=0.05-0.1, and within velocities of $0.08 \le V \le 0.09$. The spectrum from row 
T=0.07 Figure \ref{disk 60 strips},  in which the spike produced by this material lies, is shown at the top left of Figure \ref{disk part}.
 This spike  is an example of a strong effect that occurs when microlensing and flaring are combined. At T=0.25 the spike flattens, due to the flare passing over the inner edge of the disk which is near the left caustic.

\section{Conclusions}
\label{sec:conc}

In this paper we have demonstrated how the microlensing of a flare in a BELR can be modelled and analysed, and some of the effects that may be observed when a flaring BELR is  microlensed. It is clear that the behaviour will be determined by several different parameters. Those that  have  been studied previously include the source model, and the emission and velocity profiles. When a BELR is flaring  our investigations indicate that the following  will also be important:

\textbf{Strength of the flare.}
We have chosen a flare that increases the total flux by a moderate amount in the early stages ($\sim$ 50\% in the  sphere). The stronger the flare is, compared to the total BELR flux, the more noticeable it is, and the easier it will be to study as it is microlensed.

\textbf{Temporal asymmetry of the flare}.  The variation in brightness of the flare over time depends on the source model and orientation. However, making use of this would require intensive monitoring during the flare lifetime because variation can occur rapidly, particularly at the beginning of the flare, where it will be most indicative of the model. 

\textbf{Source shape with respect to the flare}. The flare in the spherical BELR is observed on the sky as an expanding and then diminishing circle. When overlaid on other source profiles the flare can break up into multiple regions, and this can produce structure in the spectrum if those regions have different velocities and different magnifications.

\textbf{High magnification regions with respect to the flare}. If the flare passes over a high magnification region of comparable size, like a caustic, this can cause a spike in the lensed spectrum. Therefore different locations will produce different microlensing behaviour due to the flare, as well as due to the source profile.

In most cases we have noticed that, at a first-order approximation,  lensing does not change the shape of the flaring BELR spectrum as the flare progresses. Sources that will exhibit some variability can be characterised as:
\begin{itemize}
\item having blue and redshifted material spatially separated on the sky,
\item not uniform, having internal structure, such as edges,
\item oriented so that the flare spreads asymmetrically over the profile over time,
\item oriented so that  the flare passes over high magnification regions of similar size to the flare.
\end{itemize}
We have found, at least at the source location of Figure \ref{locations}, that side-on orientations have produced most variability.  Q2337+0305 may therefore not be an appropriate system for microlensing studies of flaring BELRs, as \citet{poindexter2009b}  have determined the quasar is inclined by only $\sim 35\,^\circ$ (in our convention) and hence is mostly face-on. Also, Q2237+0305 will produce microlensing variability  that occurs within the duration of the flare, due to the short time-scale of events in Q2237+0305, which will complicate the analysis.

A comprehensive quantitative analysis will determine to what extent different models, and their orientations, can be differentiated. We intend to characterize the change in spectra and total magnification of the BELR, as a flare passes through it, using a small number of parameters based on simple statistics or feature extraction. Of particular importance  will also be the uncorrelated microlensing variations between the four images in Q2237+0305, which may produce the most significant means of differentiating between sources.  Since the location of the observed flare in the BELR profile may lie over caustics, and the caustic structure will not be identical at that location for each image, there will certainly be differences in the behaviour between the images. A statistical analysis will indicate how well the microlensing between the images can differentiate the BELR models during a flare, how often this may happen, what to use as an observing trigger, and what to look for in the light curves and spectra of the images during the event. 

%

\section*{Acknowledgments}

We thank the anonymous reviewer for their comments which improved the quality and content of this paper. This work was supported by the High Performance Computing Facility at The University of Sydney and the NCI National Facility at the ANU.
Hugh Garsden acknowledges  funding support from an Australian Postgraduate Award.
This work is undertaken as part of the Commonwealth Cosmology Initiative (www.thecci.org), and funded by the Australian Research Council  Discovery Project DP0665574.

\bsp

\label{lastpage}

\end{document}